\documentclass[aps,pra,twocolumn,amsmath,amssymb,superscriptaddress,floatfix]{revtex4-1}

\usepackage{graphicx}
\usepackage{multirow}
\usepackage{color}
\usepackage{ulem}
\usepackage{bm}

\renewcommand{\Vec}[1]{\bm{#1}}
\def\infinity{\infty}
\def\t#1{\textrm{#1}}
\def\ket#1{|#1\rangle }
\def\bra#1{\langle #1 |}
\def\braket#1{\langle #1 \rangle}
\newcommand{\bk}[3]{\langle {#1} | #2 |{#3} \rangle }
\def\n{\nonumber \\ }

\def\beq{\begin{equation}}
\def\eeq{\end{equation}}
\def\curv{{\boldsymbol\Omega}}
\def\k{{\bm k}}

\def\B{{\bm B}}

\def\m{{\bm m}}
\def\E{{\bm E}}

\def\dkkk{[d \bm k]}

\renewcommand{\Im}{\operatorname{Im}}

\begin{document}

\title{
Semiclassical theory of nonlinear magneto-optical responses with applications to topological Dirac/Weyl semimetals
}

\author{Takahiro Morimoto}
\affiliation{Department of Physics,
University of California, Berkeley, CA 94720}

\author{Shudan Zhong}
\affiliation{Department of Physics,
University of California, Berkeley, CA 94720}

\author{Joseph Orenstein}
\affiliation{Department of Physics,
University of California, Berkeley, CA 94720}
\affiliation{Materials Sciences Division, Lawrence Berkeley National Laboratory, Berkeley, CA 94720}

\author{Joel E. Moore}
\affiliation{Department of Physics,
University of California, Berkeley, CA 94720}
\affiliation{Materials Sciences Division, Lawrence Berkeley National Laboratory, Berkeley, CA 94720}

\date{\today}

\begin{abstract}
We study nonlinear magneto-optical responses of metals by a semiclassical Boltzmann equation approach.
We derive general formulas for linear and second order nonlinear optical effects in the presence of magnetic fields that include both Berry curvature and orbital magnetic moment.  Applied to Weyl fermions, the semiclassical approach (i) captures the directional anisotropy of linear conductivity under magnetic field as a consequence of an anisotropic $B^2$ contribution, which may explain the low-field regime of recent experiments; (ii) predicts strong second harmonic generation proportional to $B$ that is enhanced as the Fermi energy approaches the Weyl point, leading to large nonlinear Kerr rotation. Moreover, we show that the semiclassical formula for the circular photogalvanic effect arising from the Berry curvature dipole is reproduced by a full quantum calculation using a Floquet approach.
\end{abstract}

\maketitle

\section{Introduction}

The wavefunction of a single electron moving through a crystal has several geometric properties whose importance in insulators is well known.  The most celebrated example is the Berry phase derived from Bloch states.  It gives a gauge field in momentum space that underlies topological phases ranging from the integer quantum Hall effect to topological insulators.  These phases are characterized by topological invariants that can be expressed as integrals of Berry gauge fields; even in ordinary insulators, similar integrals describe important physical quantities such as electric polarization~\cite{thoulesspolarization,ksv} as well as the magnetoelectric response~\cite{qilong,essinmoorevanderbilt,essinturnermoorevanderbilt,malashevich}.

In metals, the Berry gauge field is known to give an additional term (the ``anomalous velocity'') in the semiclassical equations of motion that describe the motion in real and momentum space of a wavepacket made from Bloch states.  The anomalous velocity was originally discussed in the context of the anomalous Hall effect in magnetic metals such as iron.  The semiclassical equations can be derived systematically to linear order in applied electric and magnetic fields, under certain assumptions that we review more fully in Section II below. In several cases, such as the anomalous Hall effect~\cite{nagaosaahereview} and the gyrotropic or ``transport limit'' of the chiral magnetic effect~\cite{Zhong,Ma}, the semiclassical approach (SCA) fully reproduces the results obtained from quantum-mechanical calculations based on the Kubo formula.

The focus of this paper is the semiclassical theory of {\it nonlinear} properties of metals that are currently active subjects of experimental and theoretical investigation.  One motivation is that systematic quantum-mechanical derivations that capture all contributions to a given nonlinear order in applied fields have not as yet been achieved.  An example we consider is the chiral anomaly, which in a solid is a particular type of angle-dependent magnetoresistance with an enhanced electrical conductivity along the direction of an applied magnetic field.  This effect has been argued to exist based on linearization near isolated Dirac or Weyl singularities, but the lesson of the past few years of work on the chiral magnetic effect is that it can be dangerous to treat the singularities solely and without including all effects at a given order.  We derive a semiclassical formula for magnetotransport in the weak-field regime of this problem, and discuss that including all terms gives an answer distinct from that in other recent work, which may explain experimental observations on a Dirac semimetal in this regime~\cite{Liang_2014,Xiong_2015}.

The semiclassical equations of motion for an electron wavepacket in a metal are~\cite{Sundaram99}
\begin{subequations}
\begin{align}
\dot{\bm r}&=\frac{1}{\hbar}\bm{\nabla_k} \epsilon_{\bm k} - \dot{\bm k} \times \bm \Omega, \\
\hbar \dot{\bm k}&= - e \bm E - e \dot{\bm r} \times \bm B.
\end{align}
\label{eq: EOM}
\end{subequations}
One new contribution compared to the version in older textbooks~\cite{ashcroftmermin} is from the Berry curvature in momentum space,\begin{align}
\Omega=- \t{Im}[\bra{\nabla_{\bm k} u_{\bm k}} \times \ket{\nabla_{\bm k} u_{\bm k}}],
\end{align}
and another is from the orbital magnetic moment contribution to the energy dispersion:
$\epsilon_{\bm k}=\epsilon^0_{\bm k}-\bm{m}_{\bm k}\cdot \bm{B}$
where $H_{\bm k} \ket{u_{\bm k}}=\epsilon^0_{\bm k} \ket{u_{\bm k}}$
with $B=0$ and the orbital magnetic moment is
\begin{align}
\bm{m}_{\bm k}= -\frac{e}{2\hbar} \t{Im}[\bra{\nabla_{\bm k} u_{\bm k}} \times (H_{\bm k} -\epsilon^0_{\bm k}) \ket{\nabla_{\bm k} u_{\bm k}}].
\end{align}
(We note that we adopt the convention $e>0$.)

These equations conserve the properly defined volume in phase space and give an intuitive approach to many observable properties of metals.  However, the SCA can make erroneous predictions if used outside the regime of its validity. To illustrate this point we present, in Section II, the predictions of semiclassical and fully quantum theories of a fundamental nonlinear response in metals with low symmetry - the photogalvanic effect (PGE)~\cite{ganichevprl,diehl07,ganichevnew}. The term ``photogalvanic'' refers to the generation of a dc current by a time-varying electric field, with amplitude proportional to the square of the applied field. The PGE is distinguished from a conventional photovoltaic response by the dependence of the dc current on the polarization state of the electric field. For example, in the the circular PGE (CPGE) the direction of the dc current reverses when the polarization state of the time-varying field is changed from left to right circular. Using the SCA the CPGE has been shown to have a Berry-phase contribution~\cite{mooreorenstein} in 2D and more recently in 3D~\cite{sodemannfu} systems such as Weyl semimetals.

In Section II we show that the previous semiclassical predictions for the CPGE can be derived in a fully quantum theory by using the Floquet approach~\cite{Morimoto-Nagaosa}.  
We first derive the Berry curvature formula for CPGE in the case of two band and then generalize the derivation to the cases with many bands. This indicates that the CPGE provides a good example where the nonlinear effects that follow from semiclassical equations are exactly what is obtained from a full quantum derivation, which was previously only known in the linear case. We also show that in this same limit in which interband terms are neglected, there is close quantitative relation between CPGE and second-harmonic generation (SHG).


In sections III we derive semiclassical formulas for a variety of nonlinear effects. In particular, we systematically study nonlinear magneto-optical effects by incorporating the orbital magnetic moment, which has not been discussed previously. We show that magnetic fields modify the nonlinear Hall effect via the orbital moment of Bloch electrons.  In section IV, we apply our semiclassical formula to magneto-transport of Weyl/Dirac semimetals and study the angle-dependent magnetoresistance. We find that there exist contributions of opposite sign from orbital magnetic moment and Berry curvature in addition to the contribution of the chiral anomaly. The angular dependence that we obtain by taking into account all the contributions at the same order in the SCA is compared with recent magnetotransport experiments~\cite{Liang_2014,Xiong_2015}.  Section V applies the semiclassical formulas to nonlinear Kerr rotation (polarization rotation of SHG signals with applied magnetic fields) of Weyl semimetals. Since isotropic Weyl fermions with linear dispersion support no intraband contribution to SHG in the absence of magnetic fields, intraband contributions to SHG in such Weyl semimetals are linear in $B$, which leads to nonlinear Kerr rotation in general. We show that Weyl semimetals can exhibit giant nonlinear Kerr rotation in the infrared regime as the Fermi energy approaches to Weyl nodes. Section VI summarizes some remaining issues and open problems.

\section{Nonlinear optical effects and Floquet approach \label{sec: floquet}}
In this section, we first review formulas for the nonlinear Kerr rotation and CPGE. Previous works based on SCA showed that those nonlinear optical effects are described by a geometrical quantity, i.e., Berry curvature dipole \cite{sodemannfu}. We give an alternative derivation for those formulas based on fully quantum theoretical treatment by applying Floquet formalism for a two-band system.

\subsection{Geometrical meaning of nonlinear optics in the semiclassical approach}
In previous semiclassical works~\cite{mooreorenstein,sodemannfu}, it has been shown that the intraband contributions to SHG and CPGE have a geometrical nature that are described by Berry curvatures of Bloch wave functions.
The SHG is the second order nonlinear optical effect that is described by nonlinear current responses $\bm{j}(2\omega) e^{-2i\omega t}$ as
\begin{align}
j^{(2\omega)}_a&= \sigma_{abc}E_b E_c,
\end{align}
when the external electric field is given by
\begin{align}
\bm E(t)&= \bm E e^{-i\omega t} + \bm E^* e^{i\omega t}.
\end{align}
Nonlinear Hall effect in Ref.~\cite{sodemannfu} refers to a transverse current response that is described by $\sigma_{abb}$ with $a \neq b$. Similarly, the CPGE is the second order nonlinear optical effect in which dc photocurrent of $\bm{j}^{(0)}$ is induced by circularly polarized light as
\begin{align}
j^{(0)}_a&= \sigma_{abc}E_b E_c^*.
\end{align}
In a time reversal symmetric material, these nonlinear response tensors $\sigma$ are given by
\begin{align}
\sigma_{abc}&= \epsilon_{adc} \frac{e^3 \tau}{\hbar(1-i\omega \tau)}
\int \dkkk f_0 (\partial_b \Omega_d),
\label{eq: sigma semiclassics}
\end{align}
when the frequency $\omega$ is much smaller than the resonant frequency for optical transitions (i.e., the intraband contribution). Here, $\epsilon_{abc}$ is the totally antisymmetric tensor, $f_0$ is the Fermi distribution function, and we used the notation $\dkkk=d\bm k/(2\pi)^d$ with the dimension $d$.

We focus here on the case of a 3D material~\cite{sodemannfu} but have adopted slightly different notations for $\bm E(t)$ and $j$ from those in Ref.~\cite{sodemannfu}, which resulted in a modified expression for $\sigma$ above.  While these nonlinear effects are Fermi surface effects because one obtains $\sigma_{abc} \propto \epsilon_{adc} \int \dkkk (\partial_b f_0)  \Omega_d$ by integrating by parts, they can be also understood as currents carried by electrons in the Fermi sea with anomalous velocity originating from the Berry curvature dipole.

\begin{figure}
\begin{center}
\includegraphics[width=0.95\columnwidth]{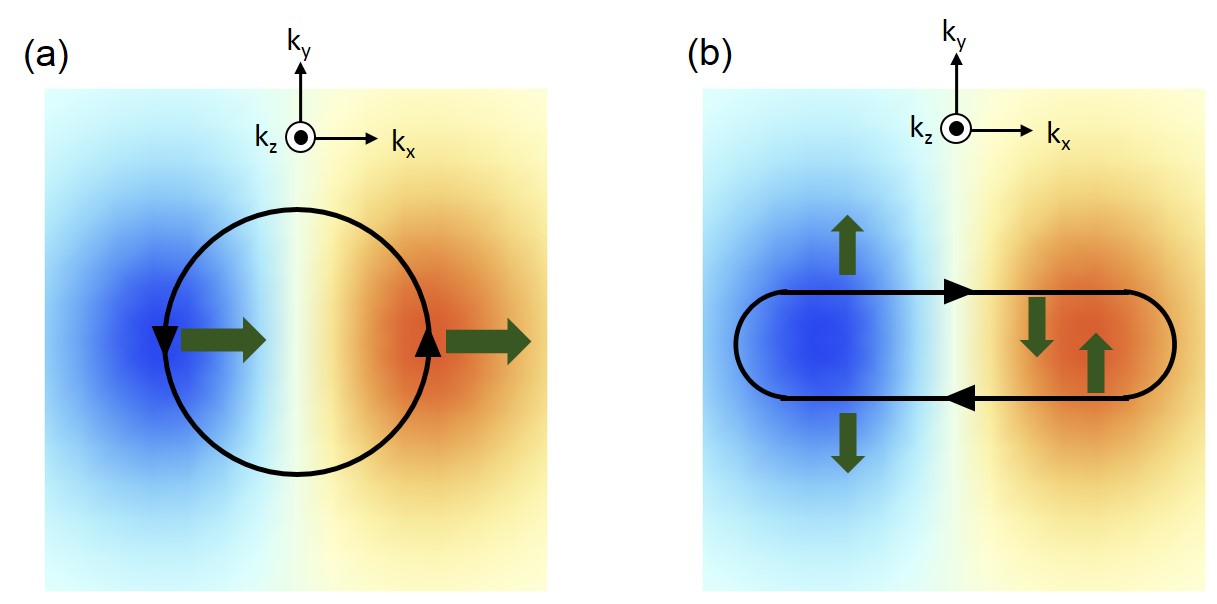}
\end{center}
\caption{ Semiclassical picture of CPGE and SHG induced by a Berry curvature dipole. The distribution of Berry curvature in momentum space is indicated by the color scale, with red region corresponding to $\Omega_z>0$ and blue region to $\Omega_z<0$. (a) CPGE arises from circular motion of the electron wave packet in momentum space driven by circularly polarized light. The dipole structure in $\Omega(k)$ induces an anomalous velocity ($\dot{\bm k} \times \bm \Omega$) in the $x$ direction denoted by green arrows. (b) SHG arises from oscillation of electron wave packet driven by linearly polarized light in the $x$ direction. The Berry curvature dipole leads to an anomalous velocity that undergoes two oscillations in the $y$ direction in one driving period. The shown configuration of Berry curvature preserves $C_{2v}$ point group symmetry (which is present for typical polar crystals that support CPGE and SHG), where the $y$-axis corresponds to the polar axis and the $yz$-plane to the mirror plane. }
\label{fig: cpge shg}
\end{figure}

The way that the anomalous velocity $(\dot{\bm k} \times \bm \Omega)$ of electron wave packets driven by an external electric field leads to CPGE and SHG is schematically illustrated in Fig.~\ref{fig: cpge shg}. Circular polarized light induces circular motion of the wave packet in momentum space [Fig.~\ref{fig: cpge shg}(a)]. In the Berry curvature dipole, the anomalous velocities in regions with $\Omega>0$ and $\Omega<0$ add, which results in dc current. Similarly, linearly polarized light induces an oscillation of wave packet as shown in Fig.~\ref{fig: cpge shg}(b). The driven wave packet exhibits anomalous velocities in the $y$ direction that oscillate twice in the driving period, which results in SHG.

\subsection{Fully quantum mechanical derivation by Floquet formalism}
Systematic derivations for the nonlinear optical effects including CPGE and SHG are presented in Sec.~\ref{sec: semiclassics nonlinear} by using SCA for general cases with finite $\bm B$. Before proceeding to general discussion with $\bm B$, we study these nonlinear optical effects from a fully quantum mechanical treatment by using a two band model. The focus of interest is whether the fully quantum mechanical expression coincides with the semiclassical formula. While SCA partially includes high energy bands through $\Omega$, it does not necessarily capture all effects of the high energy bands. Thus it is an interesting question whether the geometrical formulas for CPGE and SHG hold even in the fully quantum mechanical treatment. In the following, we study the intraband contribution to CPGE and SHG by applying the Floquet formalism and show that the Berry curvature formulas is indeed exact in the fully quantum mechanical treatment.

First we study a two band system periodically driven by an external electric field by using the Floquet formalism (for details of Floquet formalism, see Refs.~\cite{Kohler,Oka,Morimoto-Nagaosa}).
When the original Hamiltonian of the two band system is given by a Bloch Hamiltonian $H_{\t{orig}}(k)$,
the time dependent Hamiltonian of the system driven by $E(t)=E e^{-i\omega t} +E^* e^{i\omega t}$ is given by
\begin{align}
H(t,k)&=
H_{\t{orig}}(k+e A(t)), \\
A(t)&=
i \frac{E}{\omega}e^{-i\omega t} -i \frac{E^*}{\omega} e^{i\omega t},
\end{align}
which is periodic in time with $t \to t+ 2\pi/\omega$.
For such periodically driven systems, the Floquet formalism gives a concise description in terms of band picture as follows. 
The Floquet formalism is, roughly speaking, a time-direction analog of Bloch's theorem for 
time-dependent Hamiltonian $H(t)$ that satisfies $H(t+T)=H(t)$ with period $T$.
Namely, in a similar manner to Bloch's theorem, the solution for the time-periodic Schr\"odinger equation,
\begin{align}
i\hbar \frac{\partial \psi(t)}{\partial t}= H(t) \psi(t),
\end{align}
is given by a time-periodic form
\begin{align}
\psi(t)&=e^{-i \epsilon t/\hbar } \phi(t),& \phi(t+T)=\phi(t),
\end{align}
with the quasienergy $ \epsilon $.
By using the time-periodic part of the wave function $\phi(t)$, 
the time-dependent Schr\"odinger equation is rewritten as
\begin{align}
(i \hbar \partial_t +\epsilon) \phi(t) =H(t) \phi(t).
\end{align}
Since $\phi(t)$ is periodic in time, we can perform Fourier transformation of the both sides with
\begin{align}
\phi(t)=\sum_m e^{-im\omega t}\phi_m,
\end{align}
and obtain
\begin{align}
(m\hbar\omega + \epsilon)\phi_m &=\widetilde{H}_{mn} \phi_n, \\
\widetilde{H}_{mn} &= \frac{1}{T} \int_0^T dt e^{i(m-n)\omega t} H(t).
\end{align}
Here $\widetilde{H}_{mn}$ is time-independent, but has an additional matrix structure 
spanned by Floquet indices $m$ and $n$.
Thus the time-dependent Schr\"odinger equation effectively reduces to  a time-independent one in the 
Floquet formalism as,
\begin{align}
H_F \phi &= \epsilon \phi,
\end{align}
where the Floquet Hamiltonian is given by
\begin{align}
(H_F)_{mn}&= \frac{1}{T} \int_0^T dt e^{i(m-n)\omega t} H(t) 
- n \hbar\omega \delta_{mn}.
\end{align}
Floquet bands obtained by diagonalizing the Floquet Hamiltonian $H_F$ 
offer a concise understanding of the dynamics of a driven system in terms of an effective band picture.
We note that the energy spectrum of $\epsilon$ shows a periodic structure with $\hbar \omega$ as a consequence of translation symmetry with respect to the Floquet index $n$.
Thus the quasienergy spectrum is essentially described within the range $-\hbar\omega/2 \le \epsilon < \hbar\omega/2$, which is an analog of ``the first Brillouin zone'' in Bloch's theorem.

Since we consider the case of driving frequency  much lower than the band gap, we can obtain the current expectation value by studying the Floquet band that is connected to the valence band in the undriven system.
In order to do so, we use standard second order perturbation theory for
\begin{align}
H_F=H_0+H_1+H_2,
\end{align}
where $H_i$ represents a term in the Floquet Hamiltonian proportional to $A^i$.
The wave function up to the second order in $A$ reads
\begin{align}
\ket{\psi_n}&=
\ket{n}
-\sum_{m\neq n} \frac{(H_1)_{mn}}{E_m-E_n} \ket{m}
\n
&\qquad
\sum_{m\neq n}
\Bigg[
-\frac{(H_2)_{mn}}{E_m-E_n}
-\frac{(H_1)_{mn}(H_1)_{nn}}{(E_m-E_n)^2} \n
&\qquad \qquad
+\sum_{k\neq n}\frac{(H_1)_{mk}(H_1)_{kn}}{(E_m-E_n)(E_k-E_n)}
\Bigg] \ket{m}
,
\label{eq: perturbation}
\end{align}
where $H_0\ket{n}=E_n \ket{n}$.
By applying the above formula to the Floquet Hamiltonian $H_F$,
we obtain Floquet states $\ket{\psi}$ that describes the steady state under the drive of incident light.
The current responses in the steady state are obtained from perturbed Floquet states that are connected to the original valence bands. This treatment can be justified when the frequency of incident light is much smaller than the energy difference of valence and conduction bands.
(When $\omega$ satisfies conditions for optical resonances, Floquet bands originating from valence and conduction bands anticross each other. In this case, we cannot naively determine occupation of resulting Floquet bands, which requires considering the coupling to a heat bath \cite{Morimoto-Nagaosa}.)

By using the Floquet state $\ket{\psi}$ connected to the valence band,
the time dependent current in the steady state is given by
\begin{align}
J_\alpha(t)&=
\sum_{m, n} \{\t{tr}[\ket{\psi}\bra{\psi} \hat{v_\alpha}]\}_{mn}
e^{-i(m-n)\omega t},
\label{eq: j exp}
\end{align}
where  $\t{tr}$ denotes the trace over the band index,
$m,n$ are Floquet indices,
and $\hat{v_\alpha}$ is the current operator along the $\alpha$-direction is given by
\begin{align}
(\hat{v_\alpha})_{mn} &= \frac{1}{T}
\int_0^T dt e^{i(m-n)\omega t} \frac{\partial H(t)}{\partial k_\alpha}.
\end{align}
In the following, we derive representative components of nonlinear response tensor describing CPGE and SHG by using the above method.

To study CPGE we consider a system subjected to the left circularly polarized light in the $xy$ plane, where the electric field is given by
\begin{align}
\bm E(t)&=E(\bm{e_x}+i \bm{e_y}) e^{-i\omega t}+ E^* (\bm{e_x}-i \bm{e_y}) e^{i\omega t}.
\label{eq: E circular}
\end{align}
In this case, the Floquet Hamiltonian is written as
\begin{align}
H_F&=H_0+H_1, \\
(H_0)_{mn}&=
\begin{pmatrix}
\epsilon_v - n\omega & 0 \\
0& \epsilon_c - n \omega
\end{pmatrix}
\delta_{mn}, \\
(H_1)_{mn}&=
-iA^* (v_x-iv_y) \delta_{m n-1}
+iA (v_x+iv_y) \delta_{m n+1}
,
\end{align}
where $\epsilon_{v/c}$ is energy of valence/conduction band,
$v_i=\partial H_0/\partial k_i$ is the velocity operator for the static Hamiltonian, $A=E/\omega$, and we set $e=1,\hbar=1$ for simplicity.
Here we dropped the term $H_2$ proportional to $A^2$ because it does not contribute to dc photocurrent which is proportional to $AA^*$ and does not involve $A^2$ terms in the end.
Since we are interested in the second order nonlinear current responses,
it is sufficient to consider the Floquet Hamiltonian with $n=-2,\ldots,2$ by starting with the unperturbed wave function
$\ket{\psi_{ini}}=\ket{u_{v,n=0}}$.
Now we study dc current in the $x$-direction induced by circularly polarized light for the steady state described by the Floquet state in Eq.~(\ref{eq: perturbation}). The velocity operator in the $x$-direction is written up to linear order in $A$ as
\begin{align}
\hat{v_x}=
v_x\delta_{mn}
&-iA^* \partial_{k_x}(v_x - iv_y ) \delta_{m n-1}
\n
&
+iA \partial_{k_x}(v_x + iv_y ) \delta_{m n+1}
.
\end{align}
By using Eq.~(\ref{eq: j exp}),
we obtain the CPGE photocurrent $J_x=\int \dkkk j_x^{(0)}$ as
\begin{align}
j_x^{(0)}&=
\sum_{n} \{\t{tr}[\ket{\psi}\bra{\psi} \hat{v_x}]\}_{nn}
\n
&=
4\frac{|E|^2}{\omega} \Bigg\{
\frac{\t{Im}[(\partial_{k_x} v_x)_{vc}(v_y)_{cv} + (v_x)_{vc}(\partial_{k_x} v_y)_{cv}]}{(\epsilon_v-\epsilon_c)^2}
\n
&\qquad \qquad
-3
\frac{\t{Im}[(v_x)_{vc}(v_y)_{cv}][(v_x)_{vv} - (v_x)_{cc}]}{(\epsilon_v-\epsilon_c)^3}
\Bigg\},
\end{align}
where we dropped higher order terms with respect to $\omega$ by focusing on the current response in the low frequency limit.
We note that the contributions proportional to $|E|^2/\omega^2$ vanish due to the time reversal symmetry (e.g., the TRS $T=\mathcal{K}$ constrains $\t{Re}[v]$ and $\t{Im}[v]$ to be odd and even functions of $k$, respectively), which is used when going from the first line to the second line.
In the case of two band models, the Berry curvature is written as
\begin{align}
\Omega_z &=-\frac{2\t{Im}[(v_x)_{vc}(v_y)_{cv}]}{(\epsilon_v-\epsilon_c)^2},
\end{align}
and the matrix elements of $\partial_{k_i} v_j$ can be rewritten as
\begin{align}
(\partial_{k_i} v_j)_{vc} &=
\partial_{k_i}[(v_j)_{vc}] 
+ (v_j)_{vc} \left[
i (a_i)_v - i (a_i)_c \right] 
\n
&\qquad + (v_i)_{vc} \frac{(v_j)_{vv} - (v_j)_{cc}}{\epsilon_v - \epsilon_c}
,
\end{align}
with $(a_i)_{v/c} = \bra{u_{v/c}} \partial_{k_i} \ket{u_{v/c}}$.
By using these formulas, the CPGE photocurrent can be further reduced as
\begin{align}
j_x^{(0)}&=
4\frac{|E|^2}{\omega}
\frac{\partial}{\partial k_x}\left[
\frac{\t{Im}[(v_x)_{vc}(v_y)_{cv}]}{(\epsilon_v-\epsilon_c)^2}
\right]
\n
&=
-2\frac{|E|^2}{\omega}
\partial_{k_x}
\Omega_{z}.
\end{align}
The nonlinear conductivity tensor is obtained by equating the above expression and $j_x$ in terms of $\sigma$ and $\bm E(t)$ [in Eq.~(\ref{eq: E circular})] given by
\begin{align}
j_x&=-i\sigma_{xxy}|E|^2+i\sigma_{xyx}|E|^2=-2i\sigma_{xxy}|E|^2.
\end{align}
Here we used antisymmetry of imaginary part of $\sigma$ with respect to the last two indices.
This leads to
\begin{align}
\sigma_{xxy}&=\frac{1}{i\omega}
\int \dkkk \partial_{k_x} \Omega_{z},
\end{align}
and reproduces the semiclassical formula for $\sigma_{xxy}$ in Eq.~(\ref{eq: sigma semiclassics}).
We note that the factor $\tau/(1-i\omega\tau)$ in the semiclassical formula [Eq.~(\ref{eq: sigma semiclassics})] is replaced by the factor $i/\omega$ in the above formula because the $\tau\to \infinity$ limit (clean limit) is effectively taken in the Floquet perturbation theory.

Next we study SHG by using Floquet perturbation theory and the two band model in a similar manner to CPGE.
We consider a system driven by linearly polarized light in the $x$ direction as
$E_x(t)=E e^{-i\omega t}+E^* e^{i\omega t}$ and the SHG in the $y$ direction.
The corresponding Floquet Hamiltonian is given by
\begin{align}
H_F&=H_0+H_1+H_2, \\
(H_0)_{mn}&=
\begin{pmatrix}
\epsilon_v - n\omega & 0 \\
0& \epsilon_c - n \omega
\end{pmatrix}
\delta_{mn}, \\
(H_1)_{mn}&=
\left( -iA^* \delta_{m n-1}
+iA \delta_{m n+1} \right)
v_x
, \\
(H_2)_{mn}&=
\left( -\frac{(A^*)^2}{2}\delta_{m n-2}
+|A|^2 \delta_{m n}
-\frac{A^2}{2}\delta_{m n+2} \right)
\partial_{k_x} v_x
.
\end{align}
We take $\ket{\psi_{ini}}=\ket{u_{v,n=0}}$ as the unperturbed wave function and keep the part of the Floquet Hamiltonian within the range $n=-2,\ldots,2$.
The velocity operator along the $y$-direction is given by
\begin{align}
\hat{v_y}&=
v_y\delta_{mn}
+
\left( -iA^* \delta_{m n-1}
+iA \delta_{m n+1} \right)
\partial_{k_x}v_y
\n
&\qquad
+
\left( -\frac{(A^*)^2}{2}\delta_{m n-2}
+|A|^2 \delta_{m n}
-\frac{A^2}{2}\delta_{m n+2} \right)
\partial_{k_x}^2 v_y.
\end{align}
By using Eq.~(\ref{eq: j exp}),
we obtain the Fourier component of the current $J_y=\int \dkkk j_y$ proportional to $e^{-2i\omega t}$ as
\begin{align}
j_y^{(2\omega)}&=
\sum_{n} \{\t{tr}[\ket{\psi}\bra{\psi} \hat{v_y}]\}_{n+2,n}
\n
&=
-2i\frac{E^2}{i\omega}
\frac{\partial}{\partial k_x}\left[
\frac{\t{Im}[(v_x)_{vc}(v_y)_{cv}]}{(\epsilon_v-\epsilon_c)^2}
\right]
\n
&=
i\frac{E^2}{\omega}
\partial_{k_x} \Omega_{z}.
\end{align}
Here we again used the fact that the contributions proportional to $E^2/\omega^2$ vanish due to the time reversal symmetry, and also dropped contributions with higher powers of $\omega$.
The above expression indicates that the nonlinear conductivity tensor $\sigma_{yxx}$ is written as
\begin{align}
\sigma_{yxx}&=\frac{i}{\omega}
\int \dkkk \partial_{k_x} \Omega_{z}.
\end{align}
This again reproduces the semiclassical formula for $\sigma_{yxx}$ in Eq.~(\ref{eq: sigma semiclassics}).

We can extend the above analysis based on the Floquet formalism to general cases with many bands and obtain the same Berry curvature dipole formula. We sketch the derivation in the following (for details, see Appendix~\ref{app: general bands}). We consider the general Floquet Hamiltonian under the light irradiation which is given by
\begin{align}
H_F&= H_0 + H_1 + H_2, 
\end{align}
with
\begin{align}
H_1&= \sum_i  A_i v_i, &
H_2&=\frac{1}{2} \sum_{i,j} A_i A_j \partial_{k_i} v_j,
\end{align} 
where $H_0$ represents a static Hamiltonian with many bands.
By using the Floquet perturbation theory in Eq.~(\ref{eq: perturbation}) and the expression for the current in Eq.~(\ref{eq: j exp}),
we obtain the general expression for the nonlinear current response as
\begin{equation}
  \begin{aligned}
  J_r = &-  \sum_{i,j} A_iA_j \int \dkkk \\
\times &\bigg [ \sum_{n,g} [ f(\epsilon_{n}) - f(\epsilon_{g}) ]  
\left( \frac{1}{2} \frac{ (v_r)_{ng} (\partial_{k_i} v_j)_{gn}   }{\epsilon_{n} - ( \epsilon_{g} + 2\omega  ) }
+
\frac{(\partial_r v_j)_{ng} (v_i)_{gn}   }{\epsilon_{n} - ( \epsilon_{g} + \omega  ) } \right)  \\
  + & \sum_{n,g,m}   \Big( \frac{f(\epsilon_n)}{\epsilon_{n} -\epsilon_{m}- \omega } - \frac{f(\epsilon_{g})}{\epsilon_{g} -\epsilon_{m} + \omega } \Big) 
\frac{ (v_r)_{ng} (v_i)_{gm} (v_j)_{mn}   }{ \epsilon_n - (\epsilon_{g} + 2\omega  ) } 
     \\
   + & \sum_{n,g,m}  f(\epsilon_{n}) \frac{ (v_j)_{nm} (v_r)_{mg} (v_i)_{gn}  }{ (\epsilon_{n} - ( \epsilon_{g} + \omega  ) ) (\epsilon_{n} - ( \epsilon_{m} - \omega  ) )}  \\
  + & \sum_n {1 \over 2} f(\epsilon_{n}) (\partial_{k_r}\partial_{k_i}v_j)_{nn}  \bigg] , \\ 
  \end{aligned}
\end{equation}
with Fermi distribution function $f(\epsilon)$ [where $f(\epsilon_n)=1 (0)$ for occupied (unoccupied) states ].
When we expand the current $J_r$ with respect to $\omega$, the lowest order contribution in $\omega$ is proportional to $\omega A^2$ in the presence of time reversal symmetry.
In the case of many bands, the Berry curvature dipole for the $n$th band is written as
\begin{align}
\partial_{k_i} \Omega_{z,n} &= -2 \t{Im}\left[
\frac{\braket{n|\partial_{k_x}H|m}\braket{m|\partial_{k_y}H|n}}{(\epsilon_n-\epsilon_m)^2}
\right],
\end{align}
where $n$ runs over occupied bands and $m$ runs over unoccupied bands.
By using this expression for the Berry curvature dipole,
it turns out that the lowest order contribution of $J_y$ proportional to $\omega A^2$ is written as
\begin{align}
J_y&=-iw A_x^2 \int \dkkk f(\epsilon_n) \partial_{k_x} \Omega_{z,n},
\label{eq: berry curvature dipole formula, general bands}
\end{align}
which reproduces the Berry curvature dipole formula Eq.~(\ref{eq: sigma semiclassics}) for SHG in the case of many bands.
Details of the above calculation for many band cases are described in Appendix~\ref{app: general bands}. 

To summarize, we derived formulas for CPGE and SHG in the sufficiently low frequency region in a fully quantum mechanical way by using Floquet perturbation theory. This reproduces the semiclassical formula with Berry curvature dipole.

\section{Semiclassical formulas for nonlinear optical effects \label{sec: semiclassics nonlinear}}

We study nonlinear optical effects in the presence of magnetic fields using the SCA.  Deriving semiclassical formulas for nonlinear magneto-optical effects is motivated in the following senses. First, it is theoretically interesting to see how the orbital magnetic moment $m$, which is angular momentum of wave packet and also of geometrical origin, governs nonlinear optical effects and modifies previous semiclassical results for $B=0$ in Refs.~\cite{mooreorenstein,sodemannfu}.
Second, the obtained semiclassical formula for nonlinear magneto-conductivity that includes all terms proportional to $B^2 E$ is applicable to Weyl semimetals and may explain directional anisotropy of magnetoconductivity of Weyl semimetals recently reported in Ref.~\cite{Liang_2014,Xiong_2015}, which we perform in Sec.~\ref{sec: MR}.
Third, TR symmetric Weyl semimetals can support large nonlinear Kerr rotation.
Intraband contribution to SHG vanishes for $B=0$ in TR symmetric Weyl semimetals, and the SHG signal has a contribution linear in $B$. Thus application of $B$ may lead to giant nonlinear Kerr rotation.

We derive semiclassical formulas for nonlinear magneto-optical effects up to the second order in $E$.
It is convenient to rewrite the equations of motion~(\ref{eq: EOM}) to collect time derivatives on the left:
\begin{align}
\dot{\bm r}&=\frac{1}{\hbar D}[
\bm{\nabla_k} \epsilon_{\bm k} + e{\bm E} \times \bm \Omega_{\bm k} + \frac{e}{\hbar}(\bm{\nabla_k} \epsilon_{\bm k} \cdot \bm \Omega_{\bm k})\bm B], \\
\hbar \dot{\bm k}&=\frac{1}{D}[
-e{\bm E} - \frac{e}{\hbar} \bm{\nabla_k} \epsilon_{\bm k} \times \bm B - \frac{e ^2}{\hbar}(\bm{E} \cdot \bm B)\bm \Omega_{\bm k}], \\
D&=1+ \frac{e}{\hbar} \bm B \cdot \bm \Omega_{\bm k}.
\end{align}
The charge density $\rho$ and current density $\bm j$ are given by
\begin{align}
\rho&=-e\int [d\bm k] D f, \\
\bm j &=-e\int [d\bm k] (D \dot{\bm r}  + \bm{\nabla_r} \times \bm{m}_{\bm k} )f,
\end{align}
with $[d\bm k] =d\bm k/(2\pi)^3$,
where the second term of $\bm j$ is a contribution of magnetization current.
We note that the factor $D$ arises from a field-induced change of the volume of the phase space \cite{Xiao05}.
In the following, we focus on the uniform system.
In this case, the expression of the current density reduces to
\begin{align}
\bm j &=-e \int [d\bm k] [
\bm{\tilde v}_{\bm k} + \frac{e}{\hbar} \bm E \times \bm \Omega_{\bm k} + \frac{e}{\hbar}(\bm{\tilde v}_{\bm k} \cdot \bm \Omega_{\bm k})\bm B] f,
\label{eq: j with B}
\end{align}
where we used
\begin{align}
\bm{\tilde v}_{\bm p}= \bm{v}_{\bm k}-(1/\hbar)\bm{\nabla_{k}}(\bm m \cdot \bm B),
\end{align}
with $\bm{v}_{\bm k}=(1/\hbar)\bm{\nabla_{k}}\epsilon_{\bm k}^0$.

Now we focus on nonlinear responses driven by monochromatic light with the electric field $\bm E(t)=\bm{E} e^{-i\omega t}$.
We consider current responses at orders $E$, and $E^2$ as follows.
We write the distribution function in Fourier components as
\begin{align}
f&=f_0+f_1 e^{-i\omega t}+f_2 e^{-2i\omega t},
\label{eq: f0 f1 f2}
\end{align}
where $f_0$ is the unperturbed distribution function
and other terms appear in the presence of the electric field of the incident light.
The steady-state distribution function is determined by the Boltzmann equation
\begin{align}
\frac{df}{dt}=\frac{f_0-f}{\tau},
\end{align}
where
\begin{align}
\frac{df}{dt}=\bm{\dot{k}} \cdot \bm{\nabla_k} f +\partial_t f.
\label{eq: Boltzmann eq for f}
\end{align}
This gives a recursive equation for the Fourier components $f_i$.
By combining the Fourier components $f_i$ and Eq.~(\ref{eq: j with B}), we obtain nonlinear current responses in powers of $E$.
In the following, we apply the above SCA to the linear current responses and the second order nonlinear optical effects in the presence of magnetic fields.

\subsection{Linear current responses}
We first study the linear current responses with $\bm B$.
We derive the semiclassical formula for the conductivity up to the second order of $B$ in terms of Berry curvature and orbital magnetic moment.

The current response of the frequency $\omega$ is obtained from $f_1$ in Eq.~(\ref{eq: f0 f1 f2}).
By equating terms proportional to $e^{-i\omega t}$ in Eq.~(\ref{eq: Boltzmann eq for f}),
we obtain
\begin{align}
[-e{\bm E} -\frac{e ^2}{\hbar}(\bm{E} \cdot \bm B)\bm \Omega_{\bm k}] \cdot \bm{\nabla_p} f_0  - i\omega f_1
&=
-\frac{f_1}{\tau} ,
\end{align}
with $\bm{\nabla_p}=(1/\hbar)\bm{\nabla_k}$,
where we dropped the term involving $(\bm{\nabla_k} \epsilon_{\bm k}) \times \bm B$ because it is perpendicular to $\bm{\nabla_p} f_0= (1/\hbar) (\bm{\nabla_k} \epsilon_{\bm k}) \partial_\epsilon f_0$.
This leads to
\begin{align}
f_1&= \frac{-\tau}{1-i\omega\tau} \frac{1}{D}
[-e{\bm E} -\frac{e ^2}{\hbar}(\bm{E} \cdot \bm B)\bm \Omega_{\bm k}] \cdot \bm{\nabla_p} f_0
.
\end{align}
Now the current response linear in $E$ is given by
\begin{align}
\bm{j_1}&=
\frac{e \tau}{1-i\omega\tau} \int_\t{BZ} [d\bm k]
 \frac{1}{D}
\Big\{
[\bm{\tilde v}_{\bm k} + \frac{e}{\hbar}(\bm{\tilde v}_{\bm k} \cdot \bm \Omega_{\bm k})\bm B]
\n
&\qquad
\times
 [-e{\bm E} -\frac{e ^2}{\hbar}(\bm{E} \cdot \bm B)\bm \Omega_{\bm k}] \cdot \bm{\nabla_p} f_0
+ \frac{e}{\hbar} {\bm E} \times \bm \Omega_{\bm k} f_0
\Big\},
\label{eq: j1}
\end{align}
where $f_0=\theta(E_F-\epsilon_{\bm k}-\bm{m_k}\cdot \bm B)$ with the step function $\theta(x)=0 (x<0), 1 (x \ge 0)$.
This expression is reduced if we focus on the case where the electric field $\bm{E}$ is applied along the $i$th direction and the system preserves the TRS in the absence of magnetic fields.
Specifically, we consider terms up to $\propto \bm B$ that are nonvanishing with the TRS by expanding as $1/D\simeq 1+ (e/\hbar) \bm B \cdot \bm \Omega_{\bm k}$,
which leads to
\begin{align}
\bm{j_1}=
\frac{e \tau}{1-i\omega\tau}
\int_\t{BZ} [d\bm k]
&\{-\bm{v}_{\bm k} eE (\bm{v}_{\bm k})_i \partial_\epsilon f_0'
\n
&
+ \frac{e}{\hbar} (\bm E \times \bm{\Omega_k})
(\bm m \cdot \bm B) \partial_\epsilon f_0'
\},
\end{align}
with $f_0'=\theta(E_F-\epsilon_{\bm k})$, i.e., a distribution function when $\bm B=0$.
Here we used the fact that
$\partial_{p_i}, \bm{v_p}, \bm \Omega,$, and $\bm m$ are odd under the TRS.
The first term in the integral is the metallic conductivity, while the second term describes regular Hall conductivity linear in $B$ (in contrast to anomalous Hall conductivity which is nonzero in the absence of $B$).
This second term indicates that the orbital magnetic moment gives a semiclassic description related to Landau level formation in the quantum limit.
We note that there is no $B$-linear contribution to the longitudinal conductivity $\sigma_{ii}$ because the Onsager relation constrains the conductivity as $\sigma_{ij}(B)=\sigma_{ji}(-B)$ and the longitudinal conductivity should be an even function of $B$.

Next, we derive the formula for the longitudinal magnetoconductance. Its lowest order dependence on $B$ is quadratic
due to the Onsager relation. The $B^2$ contribution to the longitudinal current response is explicitly written as
\begin{widetext}
\begin{align}
\bm{j}_{B^2} =
	\frac{e^2 \tau }{\hbar}  \int_\t{BZ} [d\bm k]
\Big\{
	-\frac{e}{\hbar}\E \cdot \bm{\nabla_k} f_0(\epsilon^0) [- e (\bm{v}_{\bm{k}} \cdot \curv_\k ) ( \B \cdot \curv_\k ) \B - e \curv_\k \cdot \bm{\nabla_k} (\m \cdot \B) \B + e (\B \cdot \curv_\k)^2 \bm{v}_{\bm{k}} + (\B \cdot \curv_\k) \bm{\nabla_k}(\m \cdot \B)] \n
	+ [ \frac{1}{\hbar} \E \cdot \bm{\nabla_k}( \frac{\partial f_0(\epsilon^0)}{\partial \epsilon}  \m \cdot \B ) - \frac{e}{\hbar} (\E \cdot \B)  (  \curv_\k \cdot \bm{\nabla_k} f_0(\epsilon^0)) ] [ e( \bm{v}_{\bm{k}} \cdot \curv_\k) \B - e (\B \cdot \curv_\k) \bm{v}_{\bm{k}} - \bm{\partial_k}(\m \cdot \B) ] \n
	+ \frac{e}{\hbar} (\E \cdot \B) [\curv_\k \cdot \bm{\partial_k} ( \frac{\partial f_0(\epsilon^0)}{\partial \epsilon} \m \cdot \B )] \bm{v}_{\bm{k}} - \frac{1}{2} \E \cdot \bm{\partial_k} [ \frac{\partial^2 f_0(\epsilon^0)}{\partial \epsilon^2} ( \m \cdot \B)^2] \bm{v}_{\bm{k}}
\Big\}.
\label{eq:j1 B^2}
\end{align}
\end{widetext}
In addition to terms that contribute isotropically to the current density, there are several terms that contribute to the current density specifically along $\bm B$ which results in an anisotropic magnetoconductance if it is applied to Weyl semimetals as we discuss in Sec.~\ref{sec: MR}.

\subsection{Second order nonlinear optical effects}
We move on to the second order nonlinear optical effects which include SHG and photogalvanic effect. We derive the general formulas which will be applied to Weyl/Dirac semimetals in Sec.~\ref{sec: shg weyl}.

We consider the SHG that is described by the current response of the frequency $2\omega$.
By equating terms proportional to $e^{-2i\omega t}$ in the Boltzmann equation (\ref{eq: Boltzmann eq for f}),
we obtain
\begin{align}
\frac{1}{D}
[-e{\bm E} -\frac{e ^2}{\hbar}(\bm{E} \cdot \bm B)\bm \Omega_{\bm k}] \cdot \bm{\nabla_p} f_1 - 2i\omega f_2
&=
- \frac{f_2}{\tau} ,
\end{align}
which leads to
\begin{align}
f_2&=
\frac{\tau^2}{(1-i\omega\tau)(1-2i\omega\tau)} \frac{1}{D^2}
\n
&\quad \times
\left\{[-e{\bm E} -\frac{e ^2}{\hbar}(\bm{E} \cdot \bm B)\bm \Omega_{\bm k}] \cdot \bm{\nabla_p} \right\}^2 f_0.
\end{align}
The second order current response of the frequency $2\omega$ is given by
\begin{widetext}
\begin{align}
\bm{j_2}&=
-e \int_\t{BZ} [d\bm k]
\left\{[\bm{\tilde v}_{\bm k}  + \frac{e}{\hbar} (\bm{\tilde v}_{\bm k} \cdot \bm \Omega_{\bm k})\bm B]f_2
+ \frac{e}{\hbar} (\bm{E_0}\times \bm{\Omega}) f_1 \right\}
\n
&=
-e \int_\t{BZ} [d\bm k]
\Big\{[\bm{\tilde v}_{\bm k} + \frac{e}{\hbar} (\bm{\tilde v}_{\bm k} \cdot \bm \Omega_{\bm k})\bm B]
\frac{\tau^2}{(1-i\omega\tau)(1-2i\omega\tau)} \frac{1}{D^2}
\left\{[-e{\bm E} -\frac{e ^2}{\hbar}(\bm{E} \cdot \bm B)\bm \Omega_{\bm k}] \cdot \bm{\nabla_p} \right\}^2 f_0
\n
&\hspace{7em}
+ \frac{e}{\hbar}{\bm E} \times \bm \Omega_{\bm k}
\frac{\tau}{1-i\omega\tau} \frac{1}{D}
[-e{\bm E} -\frac{e ^2}{\hbar}(\bm{E} \cdot \bm B)\bm \Omega_{\bm k}] \cdot \bm{\nabla_p} f_0
\Big\}.
\label{eq: j2 with B}
\end{align}
\end{widetext}

Now we focus on the case of linearly polarized light where the electric field is given by $\bm E(t)=E e^{-i\omega t} \bm{e_i}$ ($\bm{e_i}$ being the unit vector along the $i$th direction),
and see how the above general formula can be simplified in several cases by assuming the TRS in the following.
First, when $\bm B=\bm 0$, Eq.~(\ref{eq: j2 with B}) reduces to
\begin{align}
\bm{j_2}(B=0)&=
\frac{-e \tau}{1-i\omega\tau}
\int_\t{BZ} [d\bm k]
\frac{e^2}{\hbar} E_i^2 (\bm{e_i} \times \partial_{p_i}\bm \Omega_{\bm k})
 f_0
.
\end{align}
This recovers the previously obtained semiclassical formula Eq.~(\ref{eq: sigma semiclassics}) for SHG.
The above expression clarifies that the transverse component of the SHG is described by the Berry curvature dipole $\partial_{p_i} \bm \Omega_{\bm k}$.
This Berry curvature dipole contribution can be nonzero only when the inversion symmetry is broken since inversion symmetry constrains $\Omega_{\bm k}=\Omega_{-\bm k}$ and causes cancellation of Berry curvature dipole between $\bm k$ and $-\bm k$~\cite{mooreorenstein,sodemannfu}.
Second, we consider the case when the magnetic field $\bm B$ is nonzero. The application of $\bm B$ leads to rotation of polarization plane of the SHG, which is known as nonlinear Kerr rotation and is an important nonlinear optical effect.
We study the nonlinear Kerr rotation by keeping contributions up to linear in $B$.
We start with the case where $\bm E$ and $\bm B$ are perpendicular to each other ($\bm{E} \cdot \bm B=0$).
The modification $\Delta \bm{j_2}$ in the first order of $B$ reads
\begin{widetext}
\begin{align}
\Delta \bm{j_2}&=
\frac{-e \tau^2}{(1-i\omega\tau)(1-2i\omega\tau)}
\int_\t{BZ} [d\bm k]
e^2 E_i^2
\left\{[
-2\bm{v_k}(\frac{e}{\hbar} \bm B\cdot \bm \Omega_{\bm k})
-\frac{1}{\hbar}\bm{\nabla_{k}}(\bm{m}\cdot \bm{B})
+\frac{e}{\hbar} (\bm{v}_{\bm k} \cdot \bm \Omega_{\bm k})\bm B
]
\partial_{p_i}^2 f_0'
- (\partial_{p_i}^2 \bm{v_k}) (\bm m \cdot \bm B) \partial_\epsilon f_0'
\right\}
.
\end{align}
\end{widetext}
Here we used $f_0=f_0'+ (\bm m \cdot \bm B) \partial_\epsilon f_0'$.
The nonlinear Kerr rotation arises from the component of $\Delta \bm{j_2}$ perpendicular to $\bm{j_2}(B=0)$ and encodes the information of
the Berry curvature $\bm \Omega$ and the orbital magnetic moment $\bm{m}$.
We note that
the term $\propto (\bm{v}_{\bm k} \cdot \bm \Omega_{\bm k})\bm B $ vanishes in the case of 2D systems (where $\bm{v}_{\bm p} \perp \bm \Omega_{\bm p}$).
Finally, we consider the case with $\bm{E} \cdot \bm B \neq 0$.
The further modification $\widetilde{\Delta \bm{j_2}}$  (in addition to $\Delta \bm{j_2}$) up to $B$ linear term is given by
\begin{align}
\widetilde{\Delta \bm{j_2}}&=
\frac{e \tau^2}{(1-i\omega\tau)(1-2i\omega\tau)} \frac{1}{D^2}
\int_\t{BZ} [d\bm k]
\frac{e^3}{\hbar} E_i^2 B_i
\bm{v}_{\bm k}
\n
&\qquad \times
[-2
\bm \Omega_{\bm k} \cdot \bm{\nabla}_{\bm p} \partial_{p_i}
- (\partial_{p_i} \bm \Omega_{\bm k})\cdot \bm{\nabla}_{\bm p}]
 f_0'
.
\end{align}

Next we derive semiclassical formula for the photogalvanic effect in the presence of $B$. The photogalvanic effect causes static dc current in the second order of $E$.
The dc component of the distribution function is also modified in the second order of $E$ as $f_0 \to f_0+\delta f_0$.
The associated Boltzmann equation is written as
\begin{align}
\frac{\delta f_0}{\tau}
&=
[-e{\bm E}^* - \frac{e ^2}{\hbar}(\bm{E}^* \cdot \bm B)\bm \Omega_{\bm k}] \cdot \bm{\nabla_p} f_1,
\end{align}
which is solved as
\begin{align}
\delta f_0
&=\frac{\tau^2}{1-i\omega\tau} [-e{\bm E}^* -\frac{e^2}{\hbar}(\bm{E}^* \cdot \bm B)\bm \Omega_{\bm k}]
\n
&\qquad \qquad \times
[-e{\bm E} -\frac{e^2}{\hbar}(\bm{E} \cdot \bm B)\bm \Omega_{\bm k}] f_0.
\end{align}
This leads to dc photovoltaic current $\delta \bm{j_0}$ given by
\begin{widetext}
\begin{align}
\delta \bm{j_0}&=
-e \int_\t{BZ} [d\bm k]
\left\{[\bm{v}_{\bm k} -\frac{1}{\hbar}\bm{\nabla_{k}}(\bm{m}\cdot \bm{B}) + \frac{e}{\hbar}(\bm{v}_{\bm k} \cdot \bm \Omega_{\bm k})\bm B]\delta f_0
+  \frac{e}{\hbar} (\bm{E_0}^*\times \bm{\Omega}) f_1 \right\}
\n
&=
-e \int_\t{BZ} [d\bm k]
\Big\{[\bm{v}_{\bm k} -\frac{1}{\hbar}\bm{\nabla_{k}}(\bm{m}\cdot \bm{B}) + \frac{e}{\hbar}(\bm{v}_{\bm k} \cdot \bm \Omega_{\bm k})\bm B]
\frac{\tau^2}{1-i\omega\tau} \frac{1}{D^2}
\left\{[-e{\bm E} -\frac{e^2}{\hbar}(\bm{E} \cdot \bm B)\bm \Omega_{\bm k}] \cdot \bm{\nabla_p} \right\}^2 f_0
\n
&\quad
+ \frac{e}{\hbar}{\bm E} \times \bm \Omega_{\bm k}
\frac{\tau}{1-i\omega\tau} \frac{1}{D}
[-e{\bm E} -e ^2(\bm{E} \cdot \bm B)\bm \Omega_{\bm k}] \cdot \bm{\nabla_p} f_0
\Big\},
\end{align}
\end{widetext}
where we write ${\bm E}^*={\bm E}$ in the second line, for simplicity.
This expression is analogous to $\bm{j_2}$ (i.e., SHG),
and indicates that the Berry curvature and the orbital magnetic moment of the Bloch bands also govern the Hall angle of dc photocurrent in the presence of an external magnetic field $\bm B$.

\begin{figure*}

\begin{center}
\includegraphics[width=1.75\columnwidth]{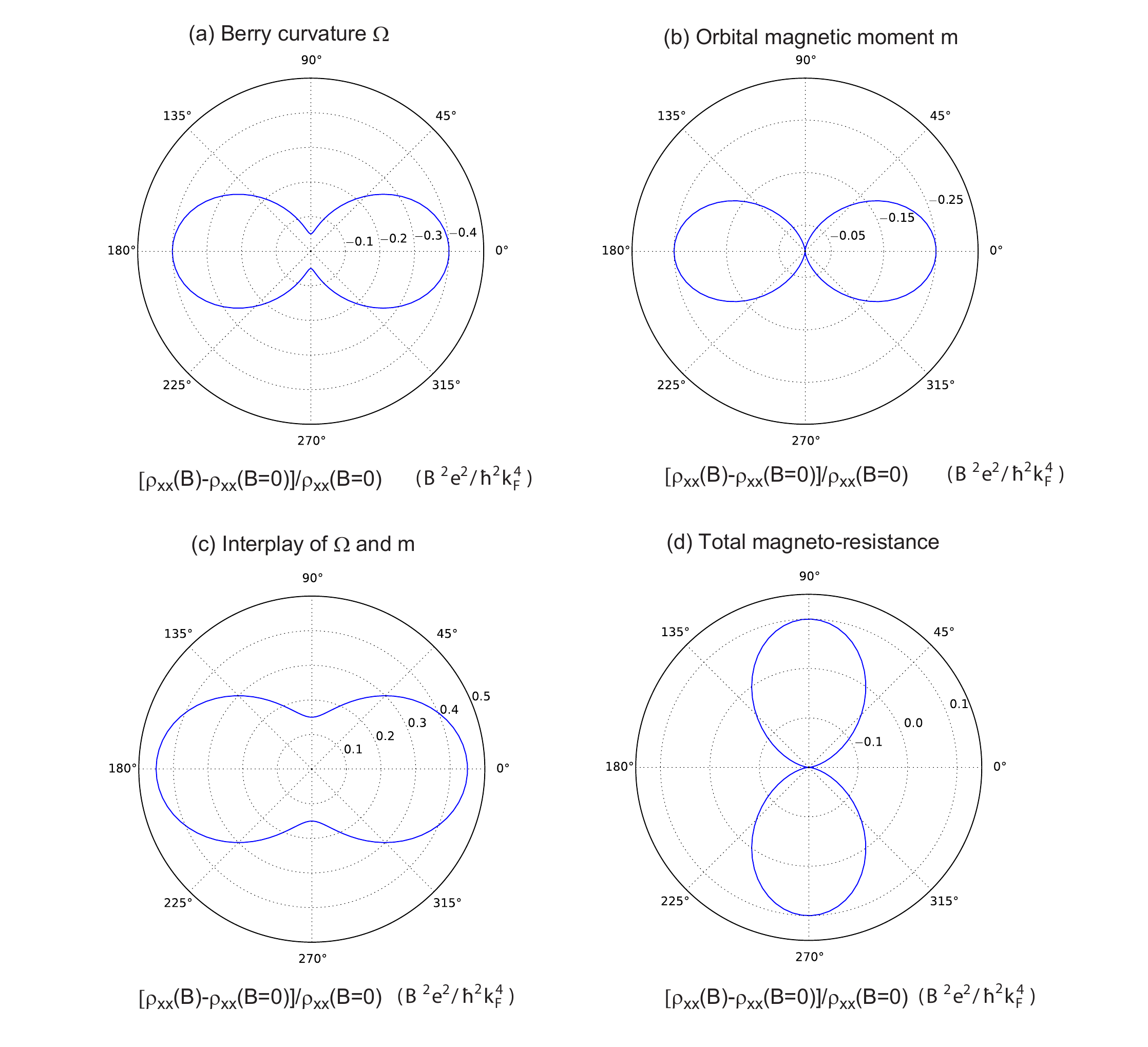}
\end{center}
\caption{ Angle-dependence of longitudinal magnetoresistance (LMR) for Weyl semimetals derived from the semiclassical approach [Eq.~(\ref{eq:j1 B^2})]. Blue lines are polar plots of the LMR $[\rho_{xx}(B)- \rho_{xx}(B=0)]/\rho_{xx}(B=0)$ as a function of the relative angle $\theta$ between $\bm{E}$ and $\bm{B}$. We show angle-dependences of contributions to the LMR from (a) the Berry curvature, (b) the orbital magnetic moment, (c) the interplay between the Berry curvature and the orbital magnetic moment, and (d) the total angle-dependence of the LMR. }
\label{fig: amr}
\end{figure*}

\section{Angle-dependent magnetoresistance \label{sec: MR}}
In this section, we study magnetoresistance by using the SCA developed in the previous section.
In particular, we focus on the current response $J\propto EB^2$ and study how the Berry curvature and the orbital magnetic moment contribute to magnetoresistance in Weyl semimetals, since the interplay of these two quantities in transport properties of Weyl semimetals has not been fully investigated except for a few studies~\cite{Zhong,Ma,Pallegrino15,Varjas16}. The obtained angle-dependence of magnetoresistance is compared with recent magneto-transport experiments for Dirac semimetals~\cite{Liang_2014,Xiong_2015}.

We consider the Hamiltonian for 
Weyl semimetals given by
\begin{align}
H=\eta v_F \bm \sigma \cdot \bm p,
\label{eq: H Weyl}
\end{align}
where $v_F$ is the Fermi velocity and $\eta= \pm 1$ specifies the chirality.
In this case, the velocity operator, the Berry curvature, the orbital magnetic moment are written as 
\begin{align}
\bm {v_k} &= v_F \bm{\hat k}, \\
\bm \Omega &= -\eta \frac{1}{2 k^2} \bm{\hat k}, \\
\bm m &=- \eta \frac{ev_F}{2 k} \bm{\hat k},
\label{eq: Omega Weyl}
\end{align}
for the conduction band, where $\hat k$ denotes the unit vector along $\bm k$.

Now we apply the semiclassics formula Eq.~(\ref{eq:j1 B^2}) for the linear current response $\bm{j_1}$ proportional to $B^2$ to Weyl semimetals and study the angle-dependent magnetoresistance.
First, we suppose that the electric field is applied in the $z$-direction as $\bm E=E \bm{e_z}$ where $\bm{e_z}$ denotes the unit vector along the $z$ direction.
In this case, the current along the $z$-direction $(j_1)_z$ is given by
\begin{subequations}
\label{eq: sigma1 weyl}
\begin{align}
(j_1)_z &=
\frac{1}{6 \pi^2 \hbar} \tau e^2 v_F k_F^2 E
+ \frac{1}{30 \pi^2 \hbar^3 k_F^2} \tau e^4 v_F B^2 E
\end{align}
when $\bm E \parallel \bm B$, and
\begin{align}
(j_1)_z &=
\frac{1}{6 \pi^2 \hbar} \tau e^2 v_F k_F^2 E
- \frac{1}{60 \pi^2 \hbar^3 k_F^2} \tau e^4 v_F B^2 E
\end{align}
\end{subequations}
when  $\bm E \perp \bm B$ (e.g. $\bm B \parallel \hat x$),
where we assumed $\tau \omega \ll 1$. Here, the first term is the isotropic dc conductivity and the second term is an anisotropic correction which originates from the $\bm E \cdot \bm B$ term related to the chiral anomaly in Weyl semimetals.
The second term accounts for the negative magnetoresistance (MR) when $\bm E \parallel \bm B$, and the positive MR when $\bm E \perp \bm B$.
Thus the semiclassical theory for the linear conductivity including effects of both $\bm \Omega$ and $\bm m$ captures the directional anisotropy of linear conductivity in the $\bm B$ field which is usually considered to be an evidence of a Weyl fermion in transport measurements.

Next, we discuss the full angle dependence of the current response in the magnetic field.
When the electric field is applied in the direction tilted by $\theta$ from the direction of the magnetic field $\bm B$,
the longitudinal magneto conductivity $\sigma(B)$ is given by
\beq
\frac{\sigma(B) -\sigma(B = 0)}{\sigma(B=0)} =  \frac{ - 1 + 3 \cos^2\theta}{10} \frac{e^2 B^2}{\hbar^2 k_F^4 }\, .
\label{eq:amr}
\eeq
Equation~(\ref{eq:amr}) does not depend on the chirality of the Weyl node nor in which band the chemical potential is located. It shows that the magnetoresistance (MR) is positive when $\E \perp \B$ and it decreases to negative as $\theta \rightarrow 0$. If we separately look at contributions to the MR from the Berry curvature and the orbital magnetic moment, we find that either the Berry curvature or the orbital magnetic moment alone gives a negative magnetoresistance (Figs.~\ref{fig: amr}a and \ref{fig: amr}b), while the interplay between the Berry curvature and the orbital magnetic moment gives a positive magnetoresistance (Fig.~\ref{fig: amr}c). As a whole, Eq.~(\ref{eq:amr}) gives the angular dependences as shown in Fig.~\ref{fig: amr}d.
We note that the anisotropic magnetoconductance in the semiclassics [Eq.~(\ref{eq:amr})] is not solely described by the contribution from the chiral anomaly.
Specifically, the contribution from the chiral anomaly which was discussed in Ref.~\cite{Son_2013} is found in the term
\begin{equation}
	\frac{-e^4 \tau}{\hbar} \int_\t{BZ} [d\bm k]  (  \curv_\k \cdot \bm{\nabla_p} f_0(\epsilon^0)) ( \bm{v}_{\bm{k}} \cdot \curv_\k)(\E \cdot \B) \B
\end{equation}
in Eq.~(\ref{eq:j1 B^2}) and gives a negative magnetoresistance in Weyl semimetals.
In contrast, there are several terms involving the orbital magnetic moment which lead to contributions of opposite signs.

A similar angular dependence of the magnetoresistance to Eq.~(\ref{eq:amr}) in the weak field region have been observed in magneto-transport experiments of Dirac semimetals \cite{Liang_2014,Xiong_2015}.
In particular, Ref.~\cite{Xiong_2015} reported that the sign change of the MR occurs around $45^\circ$ in the low $B$ region for Dirac semimetal Na$_3$Bi, which is consistent with our semiclassical result shown in Fig.~\ref{fig: amr}(d).
We note that our calculation for Weyl semimetals is also applicable to Dirac semimetals with a mild assumption that the degenerate energy bands having opposite chirality in Dirac semimetals are decoupled with each other.

Finally, we present estimates for the above nonlinear conductivities derived for Weyl semimetals.
The directional anisotropy of the linear conductivity is given by the ratio of the two terms $\propto B^0$ and $\propto B^2$ in Eq.~(\ref{eq: sigma1 weyl}).
The anisotropy ratio amounts to $0.06 (B/1\t{ T})^2$ for
typical parameters
$v_F=3\times 10^5 \t{ m/s},
E_F=10\t{ meV}
$
for the Weyl semimetal material, TaAs \cite{Weng15,Huang15}.

\section{Nonlinear magneto-optical responses in Weyl semimetals \label{sec: shg weyl}}
In this section, we study the nonlinear optical responses of Weyl semimetals in the presence of magnetic field.
Specifically, we study the second harmonic generation and the nonlinear Kerr rotation with $B$, and discuss that Weyl semimetals can support large nonlinear Kerr rotation in the infrared regime.

First, we notice that the SHG is vanishing in the absence of magnetic fields when the Weyl fermion has linear and isotropic dispersion as in Eq.~(\ref{eq: H Weyl}). The contribution from the Berry curvature dipole to the SHG cancels within the Weyl node after the $k$-integration. Thus, the application of $\bm B$ is necessary in order that the SHG is nonvanishing for isotropic Weyl fermions. In this sense, the SHG with $B$ in Weyl semimetals is a fundamental nonlinear optical effect which is related to monopole structure in the momentum space via the orbital magnetic moment.

Now we consider the SHG of Weyl fermions in the presence of the uniform magnetic field applied to the $z$-direction [$\bm B = (0,0,B)$] and study nonlinear current response proportional to $B$.
When the electric field is perpendicular to $\bm B$, e.g., $\bm E=(E,0,0)$, the nonlinear current response is given by
\begin{align}
\bm{j_2}=
\left(0,0,
\frac{e^4  v_F B}{60\pi^2 \hbar^3 \omega^2 k_F} E^2
\right),
\end{align}
where we assumed $\tau \omega \gg 1$ by focusing on the high frequency regime.
On the other hand, when the electric field is applied in the $z$-direction [$\bm E=(0,0,E)$] and is parallel to $\bm B$,
there is additional contribution to SHG from the $\bm E \cdot \bm B$ term related to the chiral anomaly of Weyl fermions.
In this case we obtain
\begin{align}
\bm{j_2}=
\left(0,0,
\frac{2e^4 v_F B}{15\pi^2 \hbar^3 \omega^2 k_F} E^2
\right).
\label{eq: sigma2 weyl}
\end{align}
This expression shows an enhancement of the SHG compared to the case of $\bm E \perp \bm B$; the chiral anomaly enhances the SHG.
Since $\bm{j_2} \propto k_F^{-1}$, the contribution to SHG proportional to $B$ becomes very large when the Fermi energy is close to the Weyl point.
This enhancement is a consequence of divergence of Berry curvature and orbital magnetic moment near the Weyl point. In this regard, the SHG of Weyl semimetals under $B$ is tied to the monopole physics in the momentum space described by Berry curvature. In practice, these divergences are cut off by the energy broadening due to the nonzero relaxation time $\tau$. This cutoff takes place around $k_F \simeq 1/(v_F \tau)$. In addition, there is another cutoff that depends on the strength of the electric field $E$. Since semiclassical treatment for Weyl fermions is only valid when $eE \tau/\hbar < k_F$ (otherwise interband effects become relevant because the shift of wavenumber exceeds the Fermi wavenumber), the divergence is also cut off around $k_F \simeq eE \tau/\hbar$.

The enhancement of SHG in Weyl semimetals can be detected as a large Kerr rotation signal.
In the case of general band structure, the SHG can become nonzero even for $B=0$ if we include the effect of band bending, e.g., by introducing a $k^2$ term in $H$. This nonzero contribution to the SHG for $B=0$ is, in general, not parallel to the above $B$-linear contribution to the SHG.
Therefore, when the magnetic field is applied,
the diverging $B$-linear contribution to SHG parallel to $\bm B$ leads to large rotation of polarization angle of SHG, and hence, large nonlinear Kerr rotation.
Incidentally, we note that when higher order terms with respect to $k$ are present in the Hamiltonian such as $k^2$ terms, additional terms having higher powers in $k_F$ arise in the current response in Eq.~(\ref{eq: sigma2 weyl}). However, when the Fermi energy is close to the Weyl point and $k_F$ is small enough, these corrections become negligible.

Finally, we estimate of magnitude of the nonlinear magneto-optical susceptibility which is given by $\chi \equiv j_2/(i\omega)\epsilon_0E^2$.
For the photon energy $\hbar\omega= 0.1\t{ eV}$ in the infrared region, the nonlinear susceptibility is estimated as
$|\chi|=1500 \times (B/1\t{ T}) \t{pm/V}$ from Eq.~(\ref{eq: sigma2 weyl}) by adopting the parameters, $v_F=3\times 10^5 \t{ m/s},$ and
$E_F=10\t{ meV}$ for Weyl semimetal material TaAs.
For comparison, GaAs, which is the representative SHG medium, shows nonlinear susceptibility of $\chi\simeq 500 \t{pm/V}$ in the visible light region~\cite{GaAs-shg}.
Thus Weyl semimetals potentially support large nonlinear Kerr rotation from the Fermi surface effect for low photon energies.
Since a recent optical measurement in TaAs reported giant SHG signals in the visible light region~\cite{Wu16}, Weyl semimetals is considered to be interesting nonlinear optical mediums in a wide range of frequency.

\section{Discussion}

We have studied CPGE and SHG in the low-frequency limit from a fully quantum mechanical treatment by using Floquet perturbation theory. By doing so, we have reproduced the expressions with Berry curvature dipole that were previously obtained from semiclassics. While we focused on second order nonlinear optical effects in this paper, the Floquet perturbation theory provides a systematic way to study general nonlinear optical responses in the low-frequency limit. Thus it will be an interesting issue to apply this method to other higher order nonlinear optical effects and investigate their geometrical meaning.

We have derived semiclassical formulas for the magneto conductance and nonlinear magneto optical effects by taking into account the orbital magnetic moment. There is an effort to partially incorporate interband effects to SCA~\cite{Gao14}. Applying this method to isotropic Weyl fermions with linear dispersion does not lead to any correction to our semiclassical formulas for magnetoconductance and SHG derived in Sec.~\ref{sec: MR} and Sec.~\ref{sec: shg weyl}. 
However, in the case of general band dispersion, the interband contributions will generate correction terms which are proportional to some inverse powers of the energy band separation. Moreover, complete formulas for these nonlinear optical effects can be derived by using a quantum mechanical treatment. The quantum treatment may be feasible for two band systems as we employed to deduce quantum formula for CPGE and SHG, while it should become very complicated in cases of a general number of bands. In particular, it will be interesting to see how the Berry curvature and orbital magnetic moment arise in the quantum mechanical treatment, as is possible for linear responses for an arbitrary number of bands~\cite{Zhong,Ma}, and what the corrections from the semiclassical formulas look like. These issues are left as future problems.

There exists another class of Weyl semimetals in which Weyl points are created by applying magnetic fields and breaking time-reversal symmetry artificially in centrosymmetric crystals. Such creation of Weyl semimetals with $B$ field was recently reported in GdPtBi~\cite{GdPtBi1,GdPtBi2}, and semiclassical analysis of magnetoresistance for those materials has been performed in Ref.~\cite{Cano16}. It would also be interesting to apply our semiclassical formula to nonlinear magneto-optical/transport properties in those field-created Weyl semimetals.

\begin{acknowledgements}
We thank M.~Kolodrubetz, B.~M.~Fregoso, and L.~Wu for fruitful discussions.
 This work
was supported by
the Gordon and Betty Moore Foundation's EPiQS Initiative Theory Center Grant (TM),
NSF DMR-1507141 (SZ),
the Gordon and Betty Moore Foundation's EPiQS Initiative through Grant GBMF4537 (JO),
and the DOE Quantum Materials program of Lawrence Berkeley National Laboratory with travel support from the Simons Foundation (JEM).
\end{acknowledgements}

\appendix

\begin{widetext}

\section{Derivation of the Berry curvature dipole formula for general bands \label{app: general bands}}
In this appendix, we apply the Floquet perturbation theory to systems with a general number of bands and derive the formula for SHG in terms of Berry curvature dipole. The derivation proceeds in a similar manner to the two band case presented in Sec.~\ref{sec: floquet}B, but with involving more band indices.

We consider the system irradiated with monochromatic light which is described by the time-dependent Hamiltonian,
\begin{equation}
  \label{eq:ham}
	\widetilde{H}(t) = H(\Vec{p} + e \Vec{A}(t) ) = H^0 + H^1 + H^2 + \cdots = H + \sum_i (\partial_{k_i} H ) e A_i e^{-i\omega t} + \sum_{i,j} \frac{1}{2}( \partial_{k_i} \partial_{k_j} H ) e^2 A_i A_j e^{-2 i\omega t} + \cdots,
\end{equation}
where $H^0\equiv H$ is the static Hamiltonian in the absence of the driving, and $\Vec{A}(t) = \Vec{A}e^{ - i\omega t}$ is the vector potential. 
For the time periodic Hamiltonian $\widetilde{H}(t)$, the Floquet Hamiltonian is defined by
\begin{align}
(H_F)_{mn}&= \frac{1}{T} \int_0^T dt e^{i(m-n)\Omega t} \widetilde{H}(t) 
- n \hbar\Omega \delta_{mn},
\end{align}
with Floquet indices $m$ an $n$. In the following, we adopt a simplified notation where we write contributions $H^i(t)$ to the Floquet Hamiltonian $H_F$ just by $H^i$.

The standard perturbation theory gives the wave function for the perturbed Floquet Hamiltonian $H_F$ as
\begin{equation}
  \label{eq: pertwave}
\ket{\psi_{\tilde n}} = \ket{\tilde n} + \sum_{\tilde g \neq \tilde n} \frac{H^1_{\tilde g \tilde n}}{\epsilon_{\tilde n} - \epsilon_{\tilde g}} \ket{\tilde g} + \sum_{\substack {\tilde n \neq \tilde m \\ \tilde g \neq \tilde n}} \left[ \frac{H^1_{\tilde g \tilde m}H^1_{\tilde m \tilde n}}{(\epsilon_{\tilde n} - \epsilon_{\tilde m})(\epsilon_{\tilde n} - \epsilon_{\tilde g})} - \frac{H^1_{\tilde n \tilde n}H^1_{\tilde g \tilde n}}{(\epsilon_{\tilde n} - \epsilon_{\tilde g})^2} + \frac{H^2_{\tilde g \tilde n}}{\epsilon_{\tilde n} - \epsilon_{\tilde g}}  \right]  \ket{\tilde g},
\end{equation}
where $\ket{\tilde n}$ is the unperturbed wave function satisfying
$H \ket{\tilde n}= \epsilon_{\tilde n} \ket{\tilde n}$,
and $\tilde n$ labels the set of the band index and the Floquet index. Here we note that $H^1_{\tilde n \tilde n}=0$ in the present case.
The explicit form of the wave function $\psi_n$ with the band index $n$ and any Floquet index (say, $0$) is given by
\begin{equation}
  \label{eq:wavefunction}
  \begin{aligned}
  \ket{\psi_n} =& \ket{n} + e \sum_{n,g} \frac{ (\partial_{k_i}
                     H A_i)_{g n}  }{\epsilon_{n} - ( \epsilon_{g} + \omega  ) } \ket{g} 
                 + {1 \over 2}e^2 \sum_{n,g} \frac{
                     ( \partial_{k_i} \partial_{k_j} H A_i A_j
                     )_{g n}  }{\epsilon_{n} - (
                     \epsilon_{g} + 2\omega  ) } \ket{g} 
  + e^2  \sum_{n,m,g} \left[ \frac{(\partial_{k_j} H A_j )_{g m} (\partial_{k_i} H A_i)_{m n}  }{(\epsilon_n - (\epsilon_{m} + \omega  )) (\epsilon_n - (\epsilon_{g} + 2\omega  )) } \right] \ket{g} \, ,
  \end{aligned}
\end{equation}
where $\ket{n}$ denotes the static wave function with the band index $n$,
$\epsilon_n$ denotes the static energy dispersion with the band index $n$,
 and $\mathcal{O}_{m,n}=\bra{m} \mathcal{O} \ket{n}$.

Now we consider the current response in the $\alpha$-direction is given by
\begin{equation}
J_\alpha (t) = -e \sum_n f(\epsilon_n) 
\sum_{m', n'} \{\t{tr}[\ket{\psi_{(n,0)}}\bra{\psi_{(n,0)}} \hat{v_\alpha}]\}_{m'n'} e^{-i(m'-n')\omega t},
\end{equation}
where $\ket{\psi_{(n,0)}}$ is the perturbed wave function with the band index $n$ and the Floquet index $0$, and $m',n'$ denote the Floquet indices.
The Fermi distribution function $f(\epsilon)$ is given by $f(\epsilon_n)=1$ for occupied bands and $f(\epsilon_n)=0$ for unoccupied bands. Since we consider the low-frequency limit where optical transition does not take place, we can assume that the occupation of the perturbed states coincides with that of the unperturbed states.
The operator $\hat{v}$ is Floquet representation of the time-dependent velocity operator $v(t)$ which is given by
\begin{align}
  \label{eq:velocity}
  (\hat{v}_i)_{m'n'} &= \frac{1}{T} \int_0^T dt e^{i(m'-n')\omega t} v_i(t) \\
 v_i(t)&= v^0_i + v^1_i + v^2_i + \cdots = \partial_{k_i} H + \sum_j (\partial_{k_i}  \partial_{k_j} H ) e A_j e^{-i\omega t} + \sum_{i,j} {1 \over 2}(\partial_{k_i} \partial_{k_j} \partial_{k_l} H ) e^2 A_j A_l e^{-2i\omega t} + \cdots \, .
\end{align}

For the real external field $\Vec{A}(t) = \Vec{A} e^{-i\omega t} + \Vec{A}e^{i\omega t}$, we obtain the second-order current response $J_r$ along the $r$-direction which is proportional to $e^{ - i2\omega t}$ as

\begin{equation}
 \begin{aligned}
	J_r  =  - & e^3 \sum_{i,j} A_i A_j \int \dkkk \sum_{n,g} \\  
\times & \bigg[   {1 \over 2}  f(\epsilon_{n})  \frac{ ( \partial_{{k}_i} \partial_{{k}_j} H )_{gn}  }{\epsilon_{n} - ( \epsilon_{g}  + 2\omega  ) } \bk{n}{\partial_{k_r}H}{g}  
 +    {1 \over 2}    f(\epsilon_{n} ) \frac{ ( \partial_{{k}_i} \partial_{{k}_j} H )_{ng}  }{\epsilon_{n} - ( \epsilon_{g}  - 2\omega  ) } \bk{g}{\partial_{k_r}H}{n}  \\
	   + &  \sum_{m}   \frac{f(\epsilon_n)}{\epsilon_{n} -\epsilon_{m}- \omega }   \frac{(\partial_{{k}_j} H )_{gm} (\partial_{{k}_i} H )_{mn}  }{ \epsilon_n - (\epsilon_{g}  + 2\omega  ) }   \bk{n}{\partial_{k_r}H}{g} 
	   +  \sum_{m}  \frac{f(\epsilon_{n})}{\epsilon_{n} -\epsilon_{m} + \omega } \frac{(\partial_{{k}_i} H )_{nm} (\partial_{{k}_j} H )_{mg}  }{ \epsilon_n - (\epsilon_{g}  - 2\omega  ) }   \bk{g}{\partial_{k_r}H}{n}  \\
   + &    f(\epsilon_{n})   \frac{ (\partial_{{k}_i} H )_{gn}  }{\epsilon_{n} - ( \epsilon_{g} + \omega  ) } \bk{n}{\partial_{k_r}\partial_{k_j} H}{g} 
   + ~   f(\epsilon_{n})   \frac{ (\partial_{{k}_i} H )_{ng}  }{\epsilon_{n} - ( \epsilon_{g} - \omega  ) } \bk{g}{\partial_{k_r}\partial_{k_j} H}{n} \\
   + &  \sum_{m}  f(\epsilon_{n}) \frac{ (\partial_{{k}_i} H )_{gn}  }{\epsilon_{n} - ( \epsilon_{g} + \omega  ) } \frac{ (\partial_{{k}_j} H )_{nm}  }{\epsilon_{n} - ( \epsilon_{m} - \omega  ) } \bk{m}{\partial_{k_r} H}{g}
    + ~ {1 \over 2}  f(\epsilon_{n}) \bk{n}{\partial_{k_r}\partial_{k_i}\partial_{k_j} H }{n} \bigg] \, . \\ 
 \end{aligned}
\end{equation}

This expression can be rewritten as
\begin{equation}
  \begin{aligned}
  J_r = - e^3 \sum_{i,j} A_iA_j \int \dkkk \sum_{n}
\bigg [ & \sum_g {1 \over 2}( f(\epsilon_{n}) - f(\epsilon_{g}) )   \frac{ \bk{n}{\partial_{k_r}H}{g} \bk{g}{\partial_{k_i} \partial_{k_j} H }{n}   }{\epsilon_{n} - ( \epsilon_{g} + 2\omega  ) }  \\
  + & \sum_{m,g}   \Big( \frac{f(\epsilon_n)}{\epsilon_{n} -\epsilon_{m}- \omega } - \frac{f(\epsilon_{g})}{\epsilon_{g} -\epsilon_{m} + \omega } \Big) 
                       \frac{ \bk{n}{\partial_{k_r}H}{g}  \bk{g}{\partial_{k_i} H }{m} \bk{m}{\partial_{k_j} H }{n}  }{ \epsilon_n - (\epsilon_{g} + 2\omega  ) } \\
  +& \sum_g   ( f(\epsilon_{n}) - f(\epsilon_{g}) ) 
   \frac{\bk{n}{\partial_{k_r}\partial_{k_j} H}{g} \bk{g}{\partial_{k_i} H }{n}  }{\epsilon_{n} - ( \epsilon_{g} + \omega  ) }  \\
   + & \sum_{m,g}  f(\epsilon_{n}) \frac{\bk{n}{\partial_{{k}_j} H }{m} \bk{m}{\partial_{k_r} H}{g} \bk{g}{\partial_{{k}_i} H }{n}  }{ (\epsilon_{n} - ( \epsilon_{g} + \omega  ) ) (\epsilon_{n} - ( \epsilon_{m} - \omega  ) )}  
  +  {1 \over 2} f(\epsilon_{n}) \bk{n}{\partial_{k_r}\partial_{k_i}\partial_{k_j} H }{n} \bigg] \, . \\ 
  \end{aligned}
\label{eq:jr}
\end{equation}

Since we are interested in the intraband effects in the low frequency limit ($\omega $ much smaller than the bandgap), we expand the current $J_r$ in terms of $\omega $ as $J_r=J_r^0 + J_r^1 + J_r^2 + \ldots$,
with $J_r^n \propto \omega^n$ . 
The lowest order term in $\omega $ is the zeroth order term which is given by
\begin{equation}
  \begin{aligned}
  J_r^0 = - & e^3 \sum_{i,j} A_iA_j \int \dkkk \sum_n \\
\times & \bigg [ \sum_{g}  {1 \over 2}( f(\epsilon_{n}) - f(\epsilon_{g}) )   \frac{ \bk{n}{\partial_{k_r}H}{g} \bk{g}{\partial_{k_i} \partial_{k_j} H }{n}   }{\epsilon_{n} -  \epsilon_{g}   }  
  +  \sum^{\prime}_{m,g}   \Big( \frac{f(\epsilon_n)}{\epsilon_{n} -\epsilon_{m}} - \frac{f(\epsilon_{g})}{\epsilon_{g} -\epsilon_{m} } \Big) 
                        \frac{ \bk{n}{\partial_{k_r}H}{g}  \bk{g}{\partial_{k_i} H }{m} \bk{m}{\partial_{k_j} H }{n}  }{ \epsilon_n - \epsilon_{g}   }  \\
  + & \sum^{\prime}_{g}   ( f(\epsilon_{n}) - f(\epsilon_{g}) ) 
   \frac{\bk{n}{\partial_{k_r}\partial_{k_j} H}{g} \bk{g}{\partial_{k_i} H }{n}  }{\epsilon_{n} -  \epsilon_{g}   }  
   +  \sum^{\prime}_{m,g}  f(\epsilon_{n}) \frac{\bk{n}{\partial_{{k}_j} H }{m} \bk{m}{\partial_{k_r} H}{g} \bk{g}{\partial_{{k}_i} H }{n}  }{ (\epsilon_{n} -  \epsilon_{g}   ) (\epsilon_{n} -  \epsilon_{m}   )}  
  \\
  - & 2 \sum^{\prime}_g   f(\epsilon_n)  \frac{ \bk{n}{\partial_{k_r}H}{g}  \bk{g}{\partial_{k_i} H }{n} \bk{n}{\partial_{k_j} H }{n}  }{ ( \epsilon_n - \epsilon_{g} )^2 }   
  - 2 \sum^{\prime}_g   f(\epsilon_g)  \frac{ \bk{n}{\partial_{k_r}H}{g}  \bk{g}{\partial_{k_i} H }{g} \bk{g}{\partial_{k_j} H }{n}  }{ ( \epsilon_n - \epsilon_{g} )^2 }   \\
  + &  \sum^{\prime}_{m}  f(\epsilon_{n}) \frac{\bk{n}{\partial_{{k}_j} H }{m} \bk{m}{\partial_{k_r} H}{n} \bk{n}{\partial_{{k}_i} H }{n}  }{  (\epsilon_{n} -  \epsilon_{m}  )^2}  
  +   \sum^{\prime}_{g} f(\epsilon_{n}) \frac{\bk{n}{\partial_{{k}_j} H }{n} \bk{n}{\partial_{k_r} H}{g} \bk{g}{\partial_{{k}_i} H }{n}  }{  (\epsilon_{n} -  \epsilon_{g}  )^2} \\
  + &  {1 \over 2} e^2 f(\epsilon_{n}) \bk{n}{\partial_{k_r}\partial_{k_i}\partial_{k_j} H }{n} \bigg] . \\ 
  \end{aligned}
\end{equation}
Here $\sum^{\prime}_g$ ($\sum^{\prime}_{m,g}$) denotes the summation where the band index $g$ ($m,g$) runs over those that do not set the denominator to zero. We note that the 5th to 8th terms are obtained by setting one energy denominator to be $1/\omega$ and expanding the other energy denominator up to $\omega^2$ in the 2nd and the 4th terms in Eq.~(\ref{eq:jr}).
In addition, the time reversal symmetry, $\mathcal{T}=K$, leads to the symmetry properties of the Hamiltonian and its eigenstates given by 
\begin{align}
H(k) &= H(-k), & \epsilon(k) &= \epsilon(-k), & \ket{n(k)} &= \bra{n(-k)}.
\end{align}
By using these properties that hold in the presence of the time-reversal symmetry, we find that the above expression for $J_r^0$ vanishes in the zeroth order. Therefore, the lowest order term is actually the first order term $J_r^1$.

The first order term in $\omega $ is written as
\begin{equation}
  \begin{aligned}
    J_r =  -  & e^3 \omega \sum_{i,j} A_iA_j   \int  \dkkk \sum_n \\ 
 \times & \bigg [  \sum^{\prime}_{g} ( f(\epsilon_{n}) - f(\epsilon_{g}) )   \frac{ \bk{n}{\partial_{k_r}H}{g} \bk{g}{\partial_{k_i} \partial_{k_j} H }{n}   }{(\epsilon_{n} - \epsilon_{g})^2 }  
  +  2 \sum^{\prime}_{m,g}\left(  \frac{f(\epsilon_n)}{\epsilon_{n} -\epsilon_{m}} - \frac{f(\epsilon_{g})}{\epsilon_{g} -\epsilon_{m} } \right) 
	  \frac{ \bk{n}{\partial_{k_r}H}{g}  \bk{g}{\partial_{k_i} H }{m} \bk{m}{\partial_{k_j} H }{n}  }{ (\epsilon_n - \epsilon_{g})^2   }  \\
	+ &  \sum^{\prime}_{m , g}\left(  \frac{f(\epsilon_n)}{(\epsilon_{n} -\epsilon_{m})^2} + \frac{f(\epsilon_{g})}{(\epsilon_{g} -\epsilon_{m})^2 } \right) 
	  \frac{ \bk{n}{\partial_{k_r}H}{g}  \bk{g}{\partial_{k_i} H }{m} \bk{m}{\partial_{k_j} H }{n}  }{ \epsilon_n - \epsilon_{g}   }  
  -  4 \sum^{\prime}_g   f(\epsilon_n)  \frac{ \bk{n}{\partial_{k_r}H}{g}  \bk{g}{\partial_{k_i} H }{n} \bk{n}{\partial_{k_j} H }{n}  }{ ( \epsilon_n - \epsilon_{g} )^3 } \\
  - & 4  \sum^{\prime}_g   f(\epsilon_g)  \frac{ \bk{n}{\partial_{k_r}H}{g}  \bk{g}{\partial_{k_i} H }{g} \bk{g}{\partial_{k_j} H }{n}  }{ ( \epsilon_n - \epsilon_{g} )^3 }  
  +   \sum^{\prime}_{g} ( f(\epsilon_{n}) - f(\epsilon_{g}) ) 
   \frac{\bk{n}{\partial_{k_r}\partial_{k_j} H}{g} \bk{g}{\partial_{k_i} H }{n}  }{(\epsilon_{n} - \epsilon_{g})^2 } \\
  -  &  \sum^{\prime}_{m}  f(\epsilon_{n}) \frac{\bk{n}{\partial_{{k}_j} H }{m} \bk{m}{\partial_{k_r} H}{n} \bk{n}{\partial_{{k}_i} H }{n}  }{  (\epsilon_{n} -  \epsilon_{m}  )^3} 
  +    \sum^{\prime}_{g}  f(\epsilon_{n}) \frac{\bk{n}{\partial_{{k}_j} H }{n} \bk{n}{\partial_{k_r} H}{g} \bk{g}{\partial_{{k}_i} H }{n}  }{  (\epsilon_{n} -  \epsilon_{g}  )^3}  \\
   - &  \sum^{\prime}_{m,g}  f(\epsilon_{n}) \frac{\bk{n}{\partial_{{k}_j} H }{m} \bk{m}{\partial_{k_r} H}{g} \bk{g}{\partial_{{k}_i} H }{n}  }{ (\epsilon_{n} -  \epsilon_{g}   ) (\epsilon_{n} -  \epsilon_{m}   )^2}  
  +   \sum^{\prime}_{m,g}  f(\epsilon_{n}) \frac{\bk{n}{\partial_{{k}_j} H }{m} \bk{m}{\partial_{k_r} H}{g} \bk{g}{\partial_{{k}_i} H }{n}  }{ (\epsilon_{n} -  \epsilon_{g}   )^2 (\epsilon_{n} -  \epsilon_{m}   )}  \bigg ] . 
   \end{aligned}
\end{equation}
By using the properties from the time reversal symmetry, this can be reduced as
\begin{equation}
  \begin{aligned}
    J_r = - & 2 e^3 \omega \sum_{i,j} A_iA_j \int \dkkk \sum_n f(\epsilon_n) \\
\times & \bigg[  \sum^{\prime}_{g}  f(\epsilon_{n})   \frac{ \bk{n}{\partial_{k_r}H}{g} \bk{g}{\partial_{k_i} \partial_{k_j} H }{n}   }{(\epsilon_{n} - \epsilon_{g})^2 }  
    +  2  \sum^{\prime}_{m,g}  \frac{1}{\epsilon_{n} -\epsilon_{m}} 
	  \frac{ \bk{n}{\partial_{k_r}H}{g}  \bk{g}{\partial_{k_i} H }{m} \bk{m}{\partial_{k_j} H }{n}  }{ (\epsilon_n - \epsilon_{g})^2   }  \\
+  &   \sum^{\prime}_{m,g}  \frac{1}{(\epsilon_{n} -\epsilon_{m})^2} 
  \frac{ \bk{n}{\partial_{k_r}H}{g}  \bk{g}{\partial_{k_i} H }{m} \bk{m}{\partial_{k_j} H }{n}  }{ \epsilon_n - \epsilon_{g}   }
  -  3  \sum^{\prime}_g    \frac{ \bk{n}{\partial_{k_r}H}{g}  \bk{g}{\partial_{k_i} H }{n} \bk{n}{\partial_{k_j} H }{n}  }{ ( \epsilon_n - \epsilon_{g} )^3 }    \\
  + &   \sum^{\prime}_g  
  \frac{\bk{n}{\partial_{k_r}\partial_{k_j} H}{g} \bk{g}{\partial_{k_i} H }{n}  }{(\epsilon_{n} - \epsilon_{g})^2 } 
  -  \sum^{\prime}_{m,g}   \frac{\bk{n}{\partial_{{k}_j} H }{m} \bk{m}{\partial_{k_r} H}{g} \bk{g}{\partial_{{k}_i} H }{n}  }{ (\epsilon_{n} -  \epsilon_{g}   ) (\epsilon_{n} -  \epsilon_{m}   )^2}  
  \bigg ] \, . 
   \end{aligned}
\end{equation}

Now let us consider the specific case relevant to the Berry curvature dipole formula. Namely we suppose that $E$ is applied along the $x$-direction and consider the current $J$ in the $y$-direction:
\begin{equation}
  \begin{aligned}
    J_y = - &2 e^3 \omega A_xA_x \int \dkkk \sum_n f(\epsilon_n) \\ 
  \times & \bigg[  \sum^{\prime}_{g}  f(\epsilon_{n})   \frac{ \bk{n}{\partial_{k_y}H}{g} \bk{g}{\partial_{k_x} \partial_{k_x} H }{n}   }{(\epsilon_{n} - \epsilon_{g})^2 }  
    +  2 \sum^{\prime}_{m,g}  \frac{1}{\epsilon_{n} -\epsilon_{m}} 
	  \frac{ \bk{n}{\partial_{k_y}H}{g}  \bk{g}{\partial_{k_x} H }{m} \bk{m}{\partial_{k_x} H }{n}  }{ (\epsilon_n - \epsilon_{g})^2   }  \\
  - & 3  \sum^{\prime}_g    \frac{ \bk{n}{\partial_{k_y}H}{g}  \bk{g}{\partial_{k_x} H }{n} \bk{n}{\partial_{k_x} H }{n}  }{ ( \epsilon_n - \epsilon_{g} )^3 }  
 	+   \sum^{\prime}_{m,g}  \frac{1}{(\epsilon_{n} -\epsilon_{m})^2} 
  \frac{ \bk{n}{\partial_{k_y}H}{g}  \bk{g}{\partial_{k_x} H }{m} \bk{m}{\partial_{k_x} H }{n}  }{ \epsilon_n - \epsilon_{g}   }  \\
  + &  \sum^{\prime}_{g}  
  \frac{\bk{n}{\partial_{k_y}\partial_{k_x} H}{g} \bk{g}{\partial_{k_x} H }{n}  }{(\epsilon_{n} - \epsilon_{g})^2 } 
  -   \sum^{\prime}_{m,g}   \frac{\bk{n}{\partial_{{k}_x} H }{m} \bk{m}{\partial_{k_y} H}{g} \bk{g}{\partial_{{k}_x} H }{n}  }{ (\epsilon_{n} -  \epsilon_{g}   ) (\epsilon_{n} -  \epsilon_{m}   )^2}  
  \bigg ] \, . 
   \end{aligned}
\end{equation}
The $\bm k$-integral of the Berry curvature dipole $\Omega_{z,n}$ for the $n$th band is explicitly written in many band systems as
\begin{equation}
  \label{curvature}
	\begin{aligned}
& \int \dkkk ~ \partial_x \Omega_{z,n}(k) =  -2 \partial_x \int \dkkk ~   \Im [\bra{\partial_x n} \partial_y n \rangle ] 
                      = i \partial_x \int \dkkk ~   \sum_{g} [\langle {\partial_x n | g } \rangle \langle {g | \partial_y n} \rangle - \langle {\partial_y n | g } \rangle \langle {g | \partial_x n} \rangle  ]
\\
                      &= i \partial_x  \int \dkkk ~   \sum^{\prime}_{g} \left[ \frac{ \bk{n}{\partial_x H}{g} \bk{g}{\partial_y H}{n} }{(\epsilon_n - \epsilon_g)^2} - \frac{ \bk{n}{\partial_y H}{g} \bk{g}{\partial_x H}{n} }{(\epsilon_n - \epsilon_g)^2} \right]\\
                      & = -2i \int \dkkk ~  \sum^{\prime}_{g} \bigg [ \frac{ \bk{n}{\partial_y H}{g} \bk{\partial_x g}{\partial_x H}{n} }{(\epsilon_n - \epsilon_g)^2} + \frac{ \bk{n}{\partial_y H}{g} \bk{g}{\partial_x \partial_x H}{n} }{(\epsilon_n - \epsilon_g)^2} +\frac{ \bk{n}{\partial_y H}{g} \bk{g}{\partial_x H}{\partial_x n} }{(\epsilon_n - \epsilon_g)^2} + \frac{ \bk{\partial_x n}{\partial_y H}{g} \bk{g}{\partial_x H}{n} }{(\epsilon_n - \epsilon_g)^2} \\
  & \qquad\qquad  +\frac{ \bk{n}{\partial_x \partial_y H}{g} \bk{g}{\partial_x H}{n} }{(\epsilon_n - \epsilon_g)^2} + \frac{ \bk{n}{\partial_y H}{\partial_x g} \bk{g}{\partial_x H}{n} }{(\epsilon_n - \epsilon_g)^2} 
   - 2 \frac{ \bk{n}{\partial_y H}{g} \bk{g}{\partial_x H}{n} }{(\epsilon_n - \epsilon_g)^3} [(v_{x})_{nn} - (v_{x})_{gg}] \bigg ]  \\
  & = -2i \int \dkkk ~  
\sum^{\prime}_{g} \bigg [
\frac{ \bk{n}{\partial_y H}{g} \bk{g}{\partial_x \partial_x H}{n} }{(\epsilon_n - \epsilon_g)^2} + \frac{ \bk{n}{\partial_x \partial_y H}{g} \bk{g}{\partial_x H}{n} }{(\epsilon_n - \epsilon_g)^2} 
- 2 \frac{ \bk{n}{\partial_y H}{g} \bk{g}{\partial_x H}{n} }{(\epsilon_n - \epsilon_g)^3} [(v_{x})_{nn} - (v_{x})_{gg}]
\bigg] \\
& \qquad\qquad + \sum^{\prime}_{g,m} \bigg [ - \frac{ \bk{n}{\partial_y H}{g} \bk{ g}{\partial_x H}{m} \bk{m}{\partial_x H}{n} }{(\epsilon_m -\epsilon_g)(\epsilon_n - \epsilon_g)^2} + \frac{ \bk{n}{\partial_y H}{g} \bk{g}{\partial_x H}{m} \bk{m}{\partial_x H}{n} }{(\epsilon_n - \epsilon_m)(\epsilon_n - \epsilon_g)^2} \\
  & \qquad\qquad \quad\quad\quad - \frac{ \bk{n}{\partial_x H}{m} \bk{m}{\partial_y H}{g} \bk{g}{\partial_x H}{n} }{(\epsilon_m - \epsilon_n)(\epsilon_n - \epsilon_g)^2} + \frac{ \bk{n}{\partial_y H}{g} \bk{g}{\partial_x H}{m} \bk{m}{\partial_x H}{n} }{(\epsilon_m - \epsilon_g)(\epsilon_n - \epsilon_m)^2} 
 \bigg ] ,
	\end{aligned}
\end{equation}
where we have used the time reversal symmetry to simplify the expressions and the equation 
$\braket{n | \partial_k m}=\bra{n} v \ket{m}/(\epsilon_m-\epsilon_n)$. We note that the region of the above $\bm{k}$-integration can be any $\mathcal{T}$-symmetric region that includes both $\bm k$ and $-\bm k$, especially, the Fermi sea satisfying $f(\epsilon_n)=1$.
	
By using Eq.(\ref{curvature}), we finally obtain
\begin{equation}
  \begin{aligned}
    J_y =  - & 2 e^3 \omega  A_xA_x  \int \dkkk \sum_n f(\epsilon_n) \\
\times & \bigg[ \frac{\partial_x \Omega_{z,n}}{-2i} 
   +   \sum^{\prime}_{g} \frac{ \bk{n}{\partial_y H}{g} \bk{ g}{\partial_x H}{n} \bk{n}{\partial_x H}{n} }{(\epsilon_n - \epsilon_g)^3}  
 	+  \sum^{\prime}_{g}  \frac{1}{(\epsilon_{n} -\epsilon_{g})^2} 
  \frac{ \bk{n}{\partial_{k_y}H}{g}  \bk{g}{\partial_{k_x} H }{g} \bk{g}{\partial_{k_x} H }{n}  }{ \epsilon_n - \epsilon_{g}   }  \\
    + & \sum^{\prime}_{g}  \frac{1}{\epsilon_{n} -\epsilon_{g}} 
	  \frac{ \bk{n}{\partial_{k_y}H}{g}  \bk{g}{\partial_{k_x} H }{g} \bk{g}{\partial_{k_x} H }{n}  }{ (\epsilon_n - \epsilon_{g})^2   }  
   -  3 \sum^{\prime}_g    \frac{ \bk{n}{\partial_{k_y}H}{g}  \bk{g}{\partial_{k_x} H }{n} \bk{n}{\partial_{k_x} H }{n}  }{ ( \epsilon_n - \epsilon_{g} )^3 }   \\
  +  &  2 \sum^{\prime}_g \frac{ \bk{n}{\partial_y H}{g} \bk{g}{\partial_x H}{n} }{(\epsilon_n - \epsilon_g)^3} [(v_{x})_{nn} - (v_{x})_{gg}] \bigg ]  \\
     = & - i e^3 \omega  A_xA_x \int \dkkk ~ \partial_x \Omega_z.   \\
  \end{aligned}
\end{equation}
This indicates that the nonlinear conductivity for the SHG is given by
\begin{align}
\sigma_{yxx}&=\frac{ie^3}{\omega}  \int \dkkk \sum_n f(\epsilon_n) \partial_{k_x} \Omega_{z,n},
\end{align}
which reproduces Eq.~(\ref{eq: berry curvature dipole formula, general bands}) in Sec.~\ref{sec: floquet}B and proves the Berry curvature dipole formula for SHG in general cases with many bands. 

\end{widetext}

\bibliography{NLOR}

\begin{thebibliography}{33}%
\makeatletter
\providecommand \@ifxundefined [1]{%
 \@ifx{#1\undefined}
}%
\providecommand \@ifnum [1]{%
 \ifnum #1\expandafter \@firstoftwo
 \else \expandafter \@secondoftwo
 \fi
}%
\providecommand \@ifx [1]{%
 \ifx #1\expandafter \@firstoftwo
 \else \expandafter \@secondoftwo
 \fi
}%
\providecommand \natexlab [1]{#1}%
\providecommand \enquote  [1]{``#1''}%
\providecommand \bibnamefont  [1]{#1}%
\providecommand \bibfnamefont [1]{#1}%
\providecommand \citenamefont [1]{#1}%
\providecommand \href@noop [0]{\@secondoftwo}%
\providecommand \href [0]{\begingroup \@sanitize@url \@href}%
\providecommand \@href[1]{\@@startlink{#1}\@@href}%
\providecommand \@@href[1]{\endgroup#1\@@endlink}%
\providecommand \@sanitize@url [0]{\catcode `\\12\catcode `\$12\catcode
  `\&12\catcode `\#12\catcode `\^12\catcode `\_12\catcode `\%12\relax}%
\providecommand \@@startlink[1]{}%
\providecommand \@@endlink[0]{}%
\providecommand \url  [0]{\begingroup\@sanitize@url \@url }%
\providecommand \@url [1]{\endgroup\@href {#1}{\urlprefix }}%
\providecommand \urlprefix  [0]{URL }%
\providecommand \Eprint [0]{\href }%
\providecommand \doibase [0]{http://dx.doi.org/}%
\providecommand \selectlanguage [0]{\@gobble}%
\providecommand \bibinfo  [0]{\@secondoftwo}%
\providecommand \bibfield  [0]{\@secondoftwo}%
\providecommand \translation [1]{[#1]}%
\providecommand \BibitemOpen [0]{}%
\providecommand \bibitemStop [0]{}%
\providecommand \bibitemNoStop [0]{.\EOS\space}%
\providecommand \EOS [0]{\spacefactor3000\relax}%
\providecommand \BibitemShut  [1]{\csname bibitem#1\endcsname}%
\let\auto@bib@innerbib\@empty
\bibitem [{\citenamefont {Thouless}(1983)}]{thoulesspolarization}%
  \BibitemOpen
  \bibfield  {author} {\bibinfo {author} {\bibfnamefont {D.~J.}\ \bibnamefont
  {Thouless}},\ }\href {\doibase 10.1103/PhysRevB.27.6083} {\bibfield
  {journal} {\bibinfo  {journal} {Phys. Rev. B}\ }\textbf {\bibinfo {volume}
  {27}},\ \bibinfo {pages} {6083} (\bibinfo {year} {1983})}\BibitemShut
  {NoStop}%
\bibitem [{\citenamefont {King-Smith}\ and\ \citenamefont
  {Vanderbilt}(1993)}]{ksv}%
  \BibitemOpen
  \bibfield  {author} {\bibinfo {author} {\bibfnamefont {R.~D.}\ \bibnamefont
  {King-Smith}}\ and\ \bibinfo {author} {\bibfnamefont {D.}~\bibnamefont
  {Vanderbilt}},\ }\href@noop {} {\bibfield  {journal} {\bibinfo  {journal}
  {Phys. Rev. B}\ }\textbf {\bibinfo {volume} {47}},\ \bibinfo {pages} {1651}
  (\bibinfo {year} {1993})}\BibitemShut {NoStop}%
\bibitem [{\citenamefont {Qi}\ \emph {et~al.}(2008)\citenamefont {Qi},
  \citenamefont {Hughes},\ and\ \citenamefont {Zhang}}]{qilong}%
  \BibitemOpen
  \bibfield  {author} {\bibinfo {author} {\bibfnamefont {X.-L.}\ \bibnamefont
  {Qi}}, \bibinfo {author} {\bibfnamefont {T.~L.}\ \bibnamefont {Hughes}}, \
  and\ \bibinfo {author} {\bibfnamefont {S.-C.}\ \bibnamefont {Zhang}},\ }\href
  {\doibase 10.1103/PhysRevB.78.195424} {\bibfield  {journal} {\bibinfo
  {journal} {Physical Review B}\ }\textbf {\bibinfo {volume} {78}},\ \bibinfo
  {eid} {195424} (\bibinfo {year} {2008})}\BibitemShut {NoStop}%
\bibitem [{\citenamefont {Essin}\ \emph {et~al.}(2009)\citenamefont {Essin},
  \citenamefont {Moore},\ and\ \citenamefont
  {Vanderbilt}}]{essinmoorevanderbilt}%
  \BibitemOpen
  \bibfield  {author} {\bibinfo {author} {\bibfnamefont {A.~M.}\ \bibnamefont
  {Essin}}, \bibinfo {author} {\bibfnamefont {J.~E.}\ \bibnamefont {Moore}}, \
  and\ \bibinfo {author} {\bibfnamefont {D.}~\bibnamefont {Vanderbilt}},\
  }\href {\doibase 10.1103/PhysRevLett.102.146805} {\bibfield  {journal}
  {\bibinfo  {journal} {Physical Review Letters}\ }\textbf {\bibinfo {volume}
  {102}},\ \bibinfo {eid} {146805} (\bibinfo {year} {2009})}\BibitemShut
  {NoStop}%
\bibitem [{\citenamefont {Essin}\ \emph {et~al.}(2010)\citenamefont {Essin},
  \citenamefont {Turner}, \citenamefont {Moore},\ and\ \citenamefont
  {Vanderbilt}}]{essinturnermoorevanderbilt}%
  \BibitemOpen
  \bibfield  {author} {\bibinfo {author} {\bibfnamefont {A.~M.}\ \bibnamefont
  {Essin}}, \bibinfo {author} {\bibfnamefont {A.~M.}\ \bibnamefont {Turner}},
  \bibinfo {author} {\bibfnamefont {J.~E.}\ \bibnamefont {Moore}}, \ and\
  \bibinfo {author} {\bibfnamefont {D.}~\bibnamefont {Vanderbilt}},\
  }\href@noop {} {\bibfield  {journal} {\bibinfo  {journal} {Physical Review
  B}\ }\textbf {\bibinfo {volume} {81}},\ \bibinfo {pages} {205104} (\bibinfo
  {year} {2010})}\BibitemShut {NoStop}%
\bibitem [{\citenamefont {Malashevich}\ \emph {et~al.}(2010)\citenamefont
  {Malashevich}, \citenamefont {Souza}, \citenamefont {Coh},\ and\
  \citenamefont {Vanderbilt}}]{malashevich}%
  \BibitemOpen
  \bibfield  {author} {\bibinfo {author} {\bibfnamefont {A.}~\bibnamefont
  {Malashevich}}, \bibinfo {author} {\bibfnamefont {I.}~\bibnamefont {Souza}},
  \bibinfo {author} {\bibfnamefont {S.}~\bibnamefont {Coh}}, \ and\ \bibinfo
  {author} {\bibfnamefont {D.}~\bibnamefont {Vanderbilt}},\ }\href@noop {}
  {\bibfield  {journal} {\bibinfo  {journal} {New Journal of Physics}\ }\textbf
  {\bibinfo {volume} {12}},\ \bibinfo {pages} {053032} (\bibinfo {year}
  {2010})}\BibitemShut {NoStop}%
\bibitem [{\citenamefont {Nagaosa}\ \emph {et~al.}(2010)\citenamefont
  {Nagaosa}, \citenamefont {Sinova}, \citenamefont {Onoda}, \citenamefont
  {MacDonald},\ and\ \citenamefont {Ong}}]{nagaosaahereview}%
  \BibitemOpen
  \bibfield  {author} {\bibinfo {author} {\bibfnamefont {N.}~\bibnamefont
  {Nagaosa}}, \bibinfo {author} {\bibfnamefont {J.}~\bibnamefont {Sinova}},
  \bibinfo {author} {\bibfnamefont {S.}~\bibnamefont {Onoda}}, \bibinfo
  {author} {\bibfnamefont {A.~H.}\ \bibnamefont {MacDonald}}, \ and\ \bibinfo
  {author} {\bibfnamefont {N.~P.}\ \bibnamefont {Ong}},\ }\href {\doibase
  10.1103/RevModPhys.82.1539} {\bibfield  {journal} {\bibinfo  {journal} {Rev.
  Mod. Phys.}\ }\textbf {\bibinfo {volume} {82}},\ \bibinfo {pages} {1539}
  (\bibinfo {year} {2010})}\BibitemShut {NoStop}%
\bibitem [{\citenamefont {Zhong}\ \emph {et~al.}(2016)\citenamefont {Zhong},
  \citenamefont {Moore},\ and\ \citenamefont {Souza}}]{Zhong}%
  \BibitemOpen
  \bibfield  {author} {\bibinfo {author} {\bibfnamefont {S.}~\bibnamefont
  {Zhong}}, \bibinfo {author} {\bibfnamefont {J.~E.}\ \bibnamefont {Moore}}, \
  and\ \bibinfo {author} {\bibfnamefont {I.}~\bibnamefont {Souza}},\ }\href
  {\doibase 10.1103/PhysRevLett.116.077201} {\bibfield  {journal} {\bibinfo
  {journal} {Phys. Rev. Lett.}\ }\textbf {\bibinfo {volume} {116}},\ \bibinfo
  {pages} {077201} (\bibinfo {year} {2016})}\BibitemShut {NoStop}%
\bibitem [{\citenamefont {Ma}\ and\ \citenamefont {Pesin}(2015)}]{Ma}%
  \BibitemOpen
  \bibfield  {author} {\bibinfo {author} {\bibfnamefont {J.}~\bibnamefont
  {Ma}}\ and\ \bibinfo {author} {\bibfnamefont {D.~A.}\ \bibnamefont {Pesin}},\
  }\href {\doibase 10.1103/PhysRevB.92.235205} {\bibfield  {journal} {\bibinfo
  {journal} {Phys. Rev. B}\ }\textbf {\bibinfo {volume} {92}},\ \bibinfo
  {pages} {235205} (\bibinfo {year} {2015})}\BibitemShut {NoStop}%
\bibitem [{\citenamefont {Liang}\ \emph {et~al.}(2014)\citenamefont {Liang},
  \citenamefont {Gibson}, \citenamefont {Ali}, \citenamefont {Liu},
  \citenamefont {Cava},\ and\ \citenamefont {Ong}}]{Liang_2014}%
  \BibitemOpen
  \bibfield  {author} {\bibinfo {author} {\bibfnamefont {T.}~\bibnamefont
  {Liang}}, \bibinfo {author} {\bibfnamefont {Q.}~\bibnamefont {Gibson}},
  \bibinfo {author} {\bibfnamefont {M.~N.}\ \bibnamefont {Ali}}, \bibinfo
  {author} {\bibfnamefont {M.}~\bibnamefont {Liu}}, \bibinfo {author}
  {\bibfnamefont {R.~J.}\ \bibnamefont {Cava}}, \ and\ \bibinfo {author}
  {\bibfnamefont {N.~P.}\ \bibnamefont {Ong}},\ }\href {\doibase
  10.1038/nmat4143} {\bibfield  {journal} {\bibinfo  {journal} {Nature
  Materials}\ }\textbf {\bibinfo {volume} {14}},\ \bibinfo {pages} {280}
  (\bibinfo {year} {2014})}\BibitemShut {NoStop}%
\bibitem [{\citenamefont {Xiong}\ \emph {et~al.}(2015)\citenamefont {Xiong},
  \citenamefont {Kushwaha}, \citenamefont {Liang}, \citenamefont {Krizan},
  \citenamefont {Hirschberger}, \citenamefont {Wang}, \citenamefont {Cava},\
  and\ \citenamefont {Ong}}]{Xiong_2015}%
  \BibitemOpen
  \bibfield  {author} {\bibinfo {author} {\bibfnamefont {J.}~\bibnamefont
  {Xiong}}, \bibinfo {author} {\bibfnamefont {S.~K.}\ \bibnamefont {Kushwaha}},
  \bibinfo {author} {\bibfnamefont {T.}~\bibnamefont {Liang}}, \bibinfo
  {author} {\bibfnamefont {J.~W.}\ \bibnamefont {Krizan}}, \bibinfo {author}
  {\bibfnamefont {M.}~\bibnamefont {Hirschberger}}, \bibinfo {author}
  {\bibfnamefont {W.}~\bibnamefont {Wang}}, \bibinfo {author} {\bibfnamefont
  {R.~J.}\ \bibnamefont {Cava}}, \ and\ \bibinfo {author} {\bibfnamefont
  {N.~P.}\ \bibnamefont {Ong}},\ }\href {\doibase 10.1126/science.aac6089}
  {\bibfield  {journal} {\bibinfo  {journal} {Science}\ }\textbf {\bibinfo
  {volume} {350}},\ \bibinfo {pages} {413} (\bibinfo {year}
  {2015})}\BibitemShut {NoStop}%
\bibitem [{\citenamefont {Sundaram}\ and\ \citenamefont
  {Niu}(1999)}]{Sundaram99}%
  \BibitemOpen
  \bibfield  {author} {\bibinfo {author} {\bibfnamefont {G.}~\bibnamefont
  {Sundaram}}\ and\ \bibinfo {author} {\bibfnamefont {Q.}~\bibnamefont {Niu}},\
  }\href {\doibase 10.1103/PhysRevB.59.14915} {\bibfield  {journal} {\bibinfo
  {journal} {Phys. Rev. B}\ }\textbf {\bibinfo {volume} {59}},\ \bibinfo
  {pages} {14915} (\bibinfo {year} {1999})}\BibitemShut {NoStop}%
\bibitem [{\citenamefont {Ashcroft}\ and\ \citenamefont
  {Mermin}(1976)}]{ashcroftmermin}%
  \BibitemOpen
  \bibfield  {author} {\bibinfo {author} {\bibfnamefont {N.}~\bibnamefont
  {Ashcroft}}\ and\ \bibinfo {author} {\bibfnamefont {N.}~\bibnamefont
  {Mermin}},\ }\href@noop {} {\emph {\bibinfo {title} {{Solid State
  Physics}}}}\ (\bibinfo  {publisher} {Saunders College},\ \bibinfo {address}
  {Philadelphia},\ \bibinfo {year} {1976})\BibitemShut {NoStop}%
\bibitem [{\citenamefont {Ganichev}\ \emph {et~al.}(2001)\citenamefont
  {Ganichev}, \citenamefont {Ivchenko}, \citenamefont {Danilov}, \citenamefont
  {Eroms}, \citenamefont {Wegscheider}, \citenamefont {Weiss},\ and\
  \citenamefont {Prettl}}]{ganichevprl}%
  \BibitemOpen
  \bibfield  {author} {\bibinfo {author} {\bibfnamefont {S.~D.}\ \bibnamefont
  {Ganichev}}, \bibinfo {author} {\bibfnamefont {E.~L.}\ \bibnamefont
  {Ivchenko}}, \bibinfo {author} {\bibfnamefont {S.~N.}\ \bibnamefont
  {Danilov}}, \bibinfo {author} {\bibfnamefont {J.}~\bibnamefont {Eroms}},
  \bibinfo {author} {\bibfnamefont {W.}~\bibnamefont {Wegscheider}}, \bibinfo
  {author} {\bibfnamefont {D.}~\bibnamefont {Weiss}}, \ and\ \bibinfo {author}
  {\bibfnamefont {W.}~\bibnamefont {Prettl}},\ }\href {\doibase
  10.1103/PhysRevLett.86.4358} {\bibfield  {journal} {\bibinfo  {journal}
  {Phys. Rev. Lett.}\ }\textbf {\bibinfo {volume} {86}},\ \bibinfo {pages}
  {4358} (\bibinfo {year} {2001})}\BibitemShut {NoStop}%
\bibitem [{\citenamefont {Diehl}\ \emph {et~al.}(2007)\citenamefont {Diehl},
  \citenamefont {Shalygin}, \citenamefont {Bel'kov}, \citenamefont {Hoffmann},
  \citenamefont {Danilov}, \citenamefont {Herrle}, \citenamefont {Tarasenko},
  \citenamefont {Schuh}, \citenamefont {Gerl}, \citenamefont {Wegscheider},
  \citenamefont {Prettl},\ and\ \citenamefont {Ganichev}}]{diehl07}%
  \BibitemOpen
  \bibfield  {author} {\bibinfo {author} {\bibfnamefont {H.}~\bibnamefont
  {Diehl}}, \bibinfo {author} {\bibfnamefont {V.~A.}\ \bibnamefont {Shalygin}},
  \bibinfo {author} {\bibfnamefont {V.~V.}\ \bibnamefont {Bel'kov}}, \bibinfo
  {author} {\bibfnamefont {C.}~\bibnamefont {Hoffmann}}, \bibinfo {author}
  {\bibfnamefont {S.~N.}\ \bibnamefont {Danilov}}, \bibinfo {author}
  {\bibfnamefont {T.}~\bibnamefont {Herrle}}, \bibinfo {author} {\bibfnamefont
  {S.~A.}\ \bibnamefont {Tarasenko}}, \bibinfo {author} {\bibfnamefont
  {D.}~\bibnamefont {Schuh}}, \bibinfo {author} {\bibfnamefont
  {C.}~\bibnamefont {Gerl}}, \bibinfo {author} {\bibfnamefont {W.}~\bibnamefont
  {Wegscheider}}, \bibinfo {author} {\bibfnamefont {W.}~\bibnamefont {Prettl}},
  \ and\ \bibinfo {author} {\bibfnamefont {S.~D.}\ \bibnamefont {Ganichev}},\
  }\href {http://stacks.iop.org/1367-2630/9/i=9/a=349} {\bibfield  {journal}
  {\bibinfo  {journal} {New J. Phys.}\ }\textbf {\bibinfo {volume} {9}},\
  \bibinfo {pages} {349} (\bibinfo {year} {2007})}\BibitemShut {NoStop}%
\bibitem [{\citenamefont {Olbrich}\ \emph {et~al.}(2009)\citenamefont
  {Olbrich}, \citenamefont {Tarasenko}, \citenamefont {Reitmaier},
  \citenamefont {Karch}, \citenamefont {Plohmann}, \citenamefont {Kvon},\ and\
  \citenamefont {Ganichev}}]{ganichevnew}%
  \BibitemOpen
  \bibfield  {author} {\bibinfo {author} {\bibfnamefont {P.}~\bibnamefont
  {Olbrich}}, \bibinfo {author} {\bibfnamefont {S.~A.}\ \bibnamefont
  {Tarasenko}}, \bibinfo {author} {\bibfnamefont {C.}~\bibnamefont
  {Reitmaier}}, \bibinfo {author} {\bibfnamefont {J.}~\bibnamefont {Karch}},
  \bibinfo {author} {\bibfnamefont {D.}~\bibnamefont {Plohmann}}, \bibinfo
  {author} {\bibfnamefont {Z.~D.}\ \bibnamefont {Kvon}}, \ and\ \bibinfo
  {author} {\bibfnamefont {S.~D.}\ \bibnamefont {Ganichev}},\ }\href {\doibase
  10.1103/PhysRevB.79.121302} {\bibfield  {journal} {\bibinfo  {journal} {Phys.
  Rev. B}\ }\textbf {\bibinfo {volume} {79}},\ \bibinfo {pages} {121302}
  (\bibinfo {year} {2009})}\BibitemShut {NoStop}%
\bibitem [{\citenamefont {Moore}\ and\ \citenamefont
  {Orenstein}(2010)}]{mooreorenstein}%
  \BibitemOpen
  \bibfield  {author} {\bibinfo {author} {\bibfnamefont {J.~E.}\ \bibnamefont
  {Moore}}\ and\ \bibinfo {author} {\bibfnamefont {J.}~\bibnamefont
  {Orenstein}},\ }\href@noop {} {\bibfield  {journal} {\bibinfo  {journal}
  {Phys. Rev. Lett.}\ }\textbf {\bibinfo {volume} {105}},\ \bibinfo {pages}
  {026805} (\bibinfo {year} {2010})}\BibitemShut {NoStop}%
\bibitem [{\citenamefont {Sodemann}\ and\ \citenamefont
  {Fu}(2015)}]{sodemannfu}%
  \BibitemOpen
  \bibfield  {author} {\bibinfo {author} {\bibfnamefont {I.}~\bibnamefont
  {Sodemann}}\ and\ \bibinfo {author} {\bibfnamefont {L.}~\bibnamefont {Fu}},\
  }\href {\doibase 10.1103/PhysRevLett.115.216806} {\bibfield  {journal}
  {\bibinfo  {journal} {Phys. Rev. Lett.}\ }\textbf {\bibinfo {volume} {115}},\
  \bibinfo {pages} {216806} (\bibinfo {year} {2015})}\BibitemShut {NoStop}%
\bibitem [{\citenamefont {Morimoto}\ and\ \citenamefont
  {Nagaosa}(2016)}]{Morimoto-Nagaosa}%
  \BibitemOpen
  \bibfield  {author} {\bibinfo {author} {\bibfnamefont {T.}~\bibnamefont
  {Morimoto}}\ and\ \bibinfo {author} {\bibfnamefont {N.}~\bibnamefont
  {Nagaosa}},\ }\href {\doibase 10.1126/sciadv.1501524} {\bibfield  {journal}
  {\bibinfo  {journal} {Science Advances}\ }\textbf {\bibinfo {volume} {2}},\
  \bibinfo {pages} {e1501524} (\bibinfo {year} {2016})}\BibitemShut {NoStop}%
\bibitem [{\citenamefont {Kohler}\ \emph {et~al.}(2005)\citenamefont {Kohler},
  \citenamefont {Lehmann},\ and\ \citenamefont {H\"{a}nggi}}]{Kohler}%
  \BibitemOpen
  \bibfield  {author} {\bibinfo {author} {\bibfnamefont {S.}~\bibnamefont
  {Kohler}}, \bibinfo {author} {\bibfnamefont {J.}~\bibnamefont {Lehmann}}, \
  and\ \bibinfo {author} {\bibfnamefont {P.}~\bibnamefont {H\"{a}nggi}},\
  }\href {\doibase http://dx.doi.org/10.1016/j.physrep.2004.11.002} {\bibfield
  {journal} {\bibinfo  {journal} {Physics Reports}\ }\textbf {\bibinfo {volume}
  {406}},\ \bibinfo {pages} {379 } (\bibinfo {year} {2005})}\BibitemShut
  {NoStop}%
\bibitem [{\citenamefont {Oka}\ and\ \citenamefont {Aoki}(2009)}]{Oka}%
  \BibitemOpen
  \bibfield  {author} {\bibinfo {author} {\bibfnamefont {T.}~\bibnamefont
  {Oka}}\ and\ \bibinfo {author} {\bibfnamefont {H.}~\bibnamefont {Aoki}},\
  }\href {\doibase 10.1103/PhysRevB.79.081406} {\bibfield  {journal} {\bibinfo
  {journal} {Phys. Rev. B}\ }\textbf {\bibinfo {volume} {79}},\ \bibinfo
  {pages} {081406} (\bibinfo {year} {2009})}\BibitemShut {NoStop}%
\bibitem [{\citenamefont {Xiao}\ \emph {et~al.}(2005)\citenamefont {Xiao},
  \citenamefont {Shi},\ and\ \citenamefont {Niu}}]{Xiao05}%
  \BibitemOpen
  \bibfield  {author} {\bibinfo {author} {\bibfnamefont {D.}~\bibnamefont
  {Xiao}}, \bibinfo {author} {\bibfnamefont {J.}~\bibnamefont {Shi}}, \ and\
  \bibinfo {author} {\bibfnamefont {Q.}~\bibnamefont {Niu}},\ }\href {\doibase
  10.1103/PhysRevLett.95.137204} {\bibfield  {journal} {\bibinfo  {journal}
  {Phys. Rev. Lett.}\ }\textbf {\bibinfo {volume} {95}},\ \bibinfo {pages}
  {137204} (\bibinfo {year} {2005})}\BibitemShut {NoStop}%
\bibitem [{\citenamefont {Pellegrino}\ \emph {et~al.}(2015)\citenamefont
  {Pellegrino}, \citenamefont {Katsnelson},\ and\ \citenamefont
  {Polini}}]{Pallegrino15}%
  \BibitemOpen
  \bibfield  {author} {\bibinfo {author} {\bibfnamefont {F.~M.~D.}\
  \bibnamefont {Pellegrino}}, \bibinfo {author} {\bibfnamefont {M.~I.}\
  \bibnamefont {Katsnelson}}, \ and\ \bibinfo {author} {\bibfnamefont
  {M.}~\bibnamefont {Polini}},\ }\href {\doibase 10.1103/PhysRevB.92.201407}
  {\bibfield  {journal} {\bibinfo  {journal} {Phys. Rev. B}\ }\textbf {\bibinfo
  {volume} {92}},\ \bibinfo {pages} {201407} (\bibinfo {year}
  {2015})}\BibitemShut {NoStop}%
\bibitem [{\citenamefont {Varjas}\ \emph {et~al.}(2016)\citenamefont {Varjas},
  \citenamefont {Grushin}, \citenamefont {Ilan},\ and\ \citenamefont
  {Moore}}]{Varjas16}%
  \BibitemOpen
  \bibfield  {author} {\bibinfo {author} {\bibfnamefont {D.}~\bibnamefont
  {Varjas}}, \bibinfo {author} {\bibfnamefont {A.~G.}\ \bibnamefont {Grushin}},
  \bibinfo {author} {\bibfnamefont {R.}~\bibnamefont {Ilan}}, \ and\ \bibinfo
  {author} {\bibfnamefont {J.~E.}\ \bibnamefont {Moore}},\ }\href@noop {}
  {\bibfield  {journal} {\bibinfo  {journal} {arXiv:1607.05278}\ } (\bibinfo
  {year} {2016})}\BibitemShut {NoStop}%
\bibitem [{\citenamefont {Son}\ and\ \citenamefont {Spivak}(2013)}]{Son_2013}%
  \BibitemOpen
  \bibfield  {author} {\bibinfo {author} {\bibfnamefont {D.~T.}\ \bibnamefont
  {Son}}\ and\ \bibinfo {author} {\bibfnamefont {B.~Z.}\ \bibnamefont
  {Spivak}},\ }\href {http://dx.doi.org/10.1103/physrevb.88.104412} {\bibfield
  {journal} {\bibinfo  {journal} {Phys. Rev. B}\ }\textbf {\bibinfo {volume}
  {88}},\ \bibinfo {pages} {104412} (\bibinfo {year} {2013})}\BibitemShut
  {NoStop}%
\bibitem [{\citenamefont {Weng}\ \emph {et~al.}(2015)\citenamefont {Weng},
  \citenamefont {Fang}, \citenamefont {Fang}, \citenamefont {Bernevig},\ and\
  \citenamefont {Dai}}]{Weng15}%
  \BibitemOpen
  \bibfield  {author} {\bibinfo {author} {\bibfnamefont {H.}~\bibnamefont
  {Weng}}, \bibinfo {author} {\bibfnamefont {C.}~\bibnamefont {Fang}}, \bibinfo
  {author} {\bibfnamefont {Z.}~\bibnamefont {Fang}}, \bibinfo {author}
  {\bibfnamefont {B.~A.}\ \bibnamefont {Bernevig}}, \ and\ \bibinfo {author}
  {\bibfnamefont {X.}~\bibnamefont {Dai}},\ }\href {\doibase
  10.1103/PhysRevX.5.011029} {\bibfield  {journal} {\bibinfo  {journal} {Phys.
  Rev. X}\ }\textbf {\bibinfo {volume} {5}},\ \bibinfo {pages} {011029}
  (\bibinfo {year} {2015})}\BibitemShut {NoStop}%
\bibitem [{\citenamefont {Huang}\ \emph {et~al.}(2015)\citenamefont {Huang},
  \citenamefont {Zhao}, \citenamefont {Long}, \citenamefont {Wang},
  \citenamefont {Chen}, \citenamefont {Yang}, \citenamefont {Liang},
  \citenamefont {Xue}, \citenamefont {Weng}, \citenamefont {Fang},
  \citenamefont {Dai},\ and\ \citenamefont {Chen}}]{Huang15}%
  \BibitemOpen
  \bibfield  {author} {\bibinfo {author} {\bibfnamefont {X.}~\bibnamefont
  {Huang}}, \bibinfo {author} {\bibfnamefont {L.}~\bibnamefont {Zhao}},
  \bibinfo {author} {\bibfnamefont {Y.}~\bibnamefont {Long}}, \bibinfo {author}
  {\bibfnamefont {P.}~\bibnamefont {Wang}}, \bibinfo {author} {\bibfnamefont
  {D.}~\bibnamefont {Chen}}, \bibinfo {author} {\bibfnamefont {Z.}~\bibnamefont
  {Yang}}, \bibinfo {author} {\bibfnamefont {H.}~\bibnamefont {Liang}},
  \bibinfo {author} {\bibfnamefont {M.}~\bibnamefont {Xue}}, \bibinfo {author}
  {\bibfnamefont {H.}~\bibnamefont {Weng}}, \bibinfo {author} {\bibfnamefont
  {Z.}~\bibnamefont {Fang}}, \bibinfo {author} {\bibfnamefont {X.}~\bibnamefont
  {Dai}}, \ and\ \bibinfo {author} {\bibfnamefont {G.}~\bibnamefont {Chen}},\
  }\href {\doibase 10.1103/PhysRevX.5.031023} {\bibfield  {journal} {\bibinfo
  {journal} {Phys. Rev. X}\ }\textbf {\bibinfo {volume} {5}},\ \bibinfo {pages}
  {031023} (\bibinfo {year} {2015})}\BibitemShut {NoStop}%
\bibitem [{\citenamefont {Bergfeld}\ and\ \citenamefont
  {Daum}(2003)}]{GaAs-shg}%
  \BibitemOpen
  \bibfield  {author} {\bibinfo {author} {\bibfnamefont {S.}~\bibnamefont
  {Bergfeld}}\ and\ \bibinfo {author} {\bibfnamefont {W.}~\bibnamefont
  {Daum}},\ }\href {\doibase 10.1103/PhysRevLett.90.036801} {\bibfield
  {journal} {\bibinfo  {journal} {Phys. Rev. Lett.}\ }\textbf {\bibinfo
  {volume} {90}},\ \bibinfo {pages} {036801} (\bibinfo {year}
  {2003})}\BibitemShut {NoStop}%
\bibitem [{\citenamefont {{Wu}}\ \emph {et~al.}(2016)\citenamefont {{Wu}},
  \citenamefont {{Patankar}}, \citenamefont {{Morimoto}}, \citenamefont
  {{Nair}}, \citenamefont {{Thewalt}}, \citenamefont {{Little}}, \citenamefont
  {{Analytis}}, \citenamefont {{Moore}},\ and\ \citenamefont
  {{Orenstein}}}]{Wu16}%
  \BibitemOpen
  \bibfield  {author} {\bibinfo {author} {\bibfnamefont {L.}~\bibnamefont
  {{Wu}}}, \bibinfo {author} {\bibfnamefont {S.}~\bibnamefont {{Patankar}}},
  \bibinfo {author} {\bibfnamefont {T.}~\bibnamefont {{Morimoto}}}, \bibinfo
  {author} {\bibfnamefont {N.~L.}\ \bibnamefont {{Nair}}}, \bibinfo {author}
  {\bibfnamefont {E.}~\bibnamefont {{Thewalt}}}, \bibinfo {author}
  {\bibfnamefont {A.}~\bibnamefont {{Little}}}, \bibinfo {author}
  {\bibfnamefont {J.~G.}\ \bibnamefont {{Analytis}}}, \bibinfo {author}
  {\bibfnamefont {J.~E.}\ \bibnamefont {{Moore}}}, \ and\ \bibinfo {author}
  {\bibfnamefont {J.}~\bibnamefont {{Orenstein}}},\ }\href@noop {} {\bibfield
  {journal} {\bibinfo  {journal} {arXiv:1609.04894}\ } (\bibinfo {year}
  {2016})}\BibitemShut {NoStop}%
\bibitem [{\citenamefont {Gao}\ \emph {et~al.}(2014)\citenamefont {Gao},
  \citenamefont {Yang},\ and\ \citenamefont {Niu}}]{Gao14}%
  \BibitemOpen
  \bibfield  {author} {\bibinfo {author} {\bibfnamefont {Y.}~\bibnamefont
  {Gao}}, \bibinfo {author} {\bibfnamefont {S.~A.}\ \bibnamefont {Yang}}, \
  and\ \bibinfo {author} {\bibfnamefont {Q.}~\bibnamefont {Niu}},\ }\href
  {\doibase 10.1103/PhysRevLett.112.166601} {\bibfield  {journal} {\bibinfo
  {journal} {Phys. Rev. Lett.}\ }\textbf {\bibinfo {volume} {112}},\ \bibinfo
  {pages} {166601} (\bibinfo {year} {2014})}\BibitemShut {NoStop}%
\bibitem [{\citenamefont {{Hirschberger}}\ \emph {et~al.}(2016)\citenamefont
  {{Hirschberger}}, \citenamefont {{Kushwaha}}, \citenamefont {{Wang}},
  \citenamefont {{Gibson}}, \citenamefont {{Belvin}}, \citenamefont
  {{Bernevig}}, \citenamefont {{Cava}},\ and\ \citenamefont {{Ong}}}]{GdPtBi1}%
  \BibitemOpen
  \bibfield  {author} {\bibinfo {author} {\bibfnamefont {M.}~\bibnamefont
  {{Hirschberger}}}, \bibinfo {author} {\bibfnamefont {S.}~\bibnamefont
  {{Kushwaha}}}, \bibinfo {author} {\bibfnamefont {Z.}~\bibnamefont {{Wang}}},
  \bibinfo {author} {\bibfnamefont {Q.}~\bibnamefont {{Gibson}}}, \bibinfo
  {author} {\bibfnamefont {C.~A.}\ \bibnamefont {{Belvin}}}, \bibinfo {author}
  {\bibfnamefont {B.~A.}\ \bibnamefont {{Bernevig}}}, \bibinfo {author}
  {\bibfnamefont {R.~J.}\ \bibnamefont {{Cava}}}, \ and\ \bibinfo {author}
  {\bibfnamefont {N.~P.}\ \bibnamefont {{Ong}}},\ }\href {\doibase
  10.1038/nmat4684} {\bibfield  {journal} {\bibinfo  {journal} {Nat. Mater.}\
  ,\ \bibinfo {pages} {doi:10.1038/nmat4684}} (\bibinfo {year}
  {2016})}\BibitemShut {NoStop}%
\bibitem [{\citenamefont {{Shekhar}}\ \emph {et~al.}(2016)\citenamefont
  {{Shekhar}}, \citenamefont {{Nayak}}, \citenamefont {{Singh}}, \citenamefont
  {{Kumar}}, \citenamefont {{Wu}}, \citenamefont {{Zhang}}, \citenamefont
  {{Komarek}}, \citenamefont {{Kampert}}, \citenamefont {{Skourski}},
  \citenamefont {{Wosnitza}}, \citenamefont {{Schnelle}}, \citenamefont
  {{McCollam}}, \citenamefont {{Zeitler}}, \citenamefont {{Kubler}},
  \citenamefont {{Parkin}}, \citenamefont {{Yan}},\ and\ \citenamefont
  {{Felser}}}]{GdPtBi2}%
  \BibitemOpen
  \bibfield  {author} {\bibinfo {author} {\bibfnamefont {C.}~\bibnamefont
  {{Shekhar}}}, \bibinfo {author} {\bibfnamefont {A.~K.}\ \bibnamefont
  {{Nayak}}}, \bibinfo {author} {\bibfnamefont {S.}~\bibnamefont {{Singh}}},
  \bibinfo {author} {\bibfnamefont {N.}~\bibnamefont {{Kumar}}}, \bibinfo
  {author} {\bibfnamefont {S.-C.}\ \bibnamefont {{Wu}}}, \bibinfo {author}
  {\bibfnamefont {Y.}~\bibnamefont {{Zhang}}}, \bibinfo {author} {\bibfnamefont
  {A.~C.}\ \bibnamefont {{Komarek}}}, \bibinfo {author} {\bibfnamefont
  {E.}~\bibnamefont {{Kampert}}}, \bibinfo {author} {\bibfnamefont
  {Y.}~\bibnamefont {{Skourski}}}, \bibinfo {author} {\bibfnamefont
  {J.}~\bibnamefont {{Wosnitza}}}, \bibinfo {author} {\bibfnamefont
  {W.}~\bibnamefont {{Schnelle}}}, \bibinfo {author} {\bibfnamefont
  {A.}~\bibnamefont {{McCollam}}}, \bibinfo {author} {\bibfnamefont
  {U.}~\bibnamefont {{Zeitler}}}, \bibinfo {author} {\bibfnamefont
  {J.}~\bibnamefont {{Kubler}}}, \bibinfo {author} {\bibfnamefont {S.~S.~P.}\
  \bibnamefont {{Parkin}}}, \bibinfo {author} {\bibfnamefont {B.}~\bibnamefont
  {{Yan}}}, \ and\ \bibinfo {author} {\bibfnamefont {C.}~\bibnamefont
  {{Felser}}},\ }\href@noop {} {\bibfield  {journal} {\bibinfo  {journal}
  {arXiv:1604.01641}\ } (\bibinfo {year} {2016})}\BibitemShut {NoStop}%
\bibitem [{\citenamefont {{Cano}}\ \emph {et~al.}(2016)\citenamefont {{Cano}},
  \citenamefont {{Bradlyn}}, \citenamefont {{Wang}}, \citenamefont
  {{Hirschberger}}, \citenamefont {{Ong}},\ and\ \citenamefont
  {{Bernevig}}}]{Cano16}%
  \BibitemOpen
  \bibfield  {author} {\bibinfo {author} {\bibfnamefont {J.}~\bibnamefont
  {{Cano}}}, \bibinfo {author} {\bibfnamefont {B.}~\bibnamefont {{Bradlyn}}},
  \bibinfo {author} {\bibfnamefont {Z.}~\bibnamefont {{Wang}}}, \bibinfo
  {author} {\bibfnamefont {M.}~\bibnamefont {{Hirschberger}}}, \bibinfo
  {author} {\bibfnamefont {N.~P.}\ \bibnamefont {{Ong}}}, \ and\ \bibinfo
  {author} {\bibfnamefont {B.~A.}\ \bibnamefont {{Bernevig}}},\ }\href@noop {}
  {\bibfield  {journal} {\bibinfo  {journal} {arXiv:1604.08601}\ } (\bibinfo
  {year} {2016})}\BibitemShut {NoStop}%
\end{thebibliography}%

\end{document}